\documentclass[journal]{IEEEtran}
\usepackage{epstopdf}
\usepackage{setspace}
\usepackage{amsmath}
\usepackage{amssymb}
\usepackage{stfloats}
\usepackage{amsthm}
\usepackage{multirow}
\usepackage[mathscr]{euscript}
\usepackage{amsmath,amssymb,amsfonts}
\usepackage{amsthm}
\usepackage{bbold}
\usepackage{color}
\usepackage{graphicx}
\usepackage{subfigure}  
\usepackage[]{algorithmicx}
\usepackage{algpseudocode,algorithm}
\usepackage{cite}
\usepackage{mathtools}
\usepackage{stmaryrd}
\usepackage[mathscr]{euscript}
\usepackage{array}
\usepackage{mathrsfs}

\usepackage{soul}

\newtheorem{theorem}{Theorem}
\newtheorem{lemma}{Lemma}

\newtheorem{definition}{Definition}
\usepackage{authblk}
\usepackage{dsfont}
\usepackage{amsfonts}
\usepackage{amssymb}
\usepackage{amsmath}
\usepackage[subtle,tracking = normal, charwidths = tight]{savetrees}

\DeclareMathOperator{\sinc}{sinc}
\DeclareMathOperator{\EX}{\mathop{{}\mathbb{E}}}
\def\BibTeX{{\rm B\kern-.05em{\sc i\kern-.025em b}\kern-.08em
    T\kern-.1667em\lower.7ex\hbox{E}\kern-.125emX}}
\DeclareMathOperator{\Var}{{Var}}

\author{   {Hila~Naaman, {\it Student Member, IEEE},  Neil~Irwin~Bernardo, {\it Student Member, IEEE}, Alejandro~Cohen, {\it Member, IEEE}, Yonina C. Eldar, {\it Fellow, IEEE}}

	\thanks{\scriptsize H. Naaman and Y. C. Eldar are with the Faculty of Math and Computer Science, Weizmann Institute of Science, Israel. N.I. Bernardo is with the Electrical and Electronics Engineering Institute, University of the Philippines Diliman, Quezon City 1101, Philippines. A. Cohen is with the Faculty of Electrical and Computer Engineering, Technion, Israel Institute of Technology. Emails: hila.naaman@weizmann.ac.il, neil .bernardo@eee.upd.edu.ph, alecohen@technion.ac.il, yonina.eldar@weizmann.ac.il}
\thanks{Parts of this work were presented at the European Signal Processing Conference, EUSIPCO, September 2022 \cite{naaman2021time}.\

 This research was partially supported by the European Union’s Horizon 2020 research and innovation program under grant No. 101000967-ERC-CoDeS, by the Israel Science Foundation under grant no. 0100101, and by the QuantERA grant  C’MON-QSENS.}
}

\makeatletter
\let\NAT@parse\undefined
\makeatother
\usepackage[colorlinks,citecolor=green,urlcolor=blue,bookmarks=false,hypertexnames=true]{hyperref}  

\begin{document}

\title{Time Encoding Quantization of Bandlimited \\ and Finite-Rate-of-Innovation Signals}
\maketitle
\begin{abstract}

This paper studies the impact of quantization in integrate-and-fire time encoding machine (IF-TEM) sampler used for bandlimited (BL) and finite-rate-of-innovation (FRI) signals. An upper bound is derived for the mean squared error (MSE) of IF-TEM sampler and is compared against that of classical analog-to-digital converters (ADCs) with uniform sampling and quantization. The interplay between a signal's energy, bandwidth, and peak amplitude is used to identify how the MSE of IF-TEM sampler with quantization is influenced by these parameters. More precisely, the quantization step size of the IF-TEM sampler can be reduced when the maximum frequency of a bandlimited signal or the number of pulses of an FRI signal is increased. Leveraging this insight, specific parameter settings are identified for which the quantized IF-TEM sampler achieves an MSE bound that is roughly 8 dB lower than that of a classical ADC with the same number of bits. Experimental results validate the theoretical conclusions.

\end{abstract}
\begin{IEEEkeywords}
Quantization, time encoding machine, bandlimited signals, integrate-and-fire, finite-rate-of-innovation
\end{IEEEkeywords}
\section{Introduction}

Sampling and quantization are fundamental topics in signal processing and play a crucial role in numerous applications such as telecommunications, radar, biomedical and robotic systems \cite{vaseghi2008advanced,berger1998lossy,fang2021cmos}. Traditionally, bandlimited (BL) signals are sampled at a rate greater than or equal to that dictated by the Nyquist-Shannon sampling theorem, where the samples are acquired in discrete, equally spaced intervals along time \cite{nyquist1928certain,eldar2015sampling}. These sampling systems are almost exclusively synchronized to a global clock. Generally, such clocks are power-hungry, expensive, and susceptible to electromagnetic interference at high sampling rates, which incur high engineering costs, particularly in the context of deep submicron very-large-scale integration (VLSI) \cite{miskowicz2006dynamic,koscielnik2007designing,naaman2022uniqueness,mulleti2023power,mulleti2024power}. Therefore, asynchronous circuit systems and architectures that eliminate the need for a global clock have been proposed to achieve more energy-efficient designs, immunity to metastable behavior, and reduced electromagnetic interference \cite{miskowicz2018event,1199179}.

An integrate-and-fire time encoding machine (IF-TEM) is a popular asynchronous sampler due to its low power consumption \cite{roza1997analog} and simple hardware design \cite{feichtinger2012approximate}. In this mechanism, the non-negative input signal is first integrated and then compared to a threshold. The differences between successive time instances at which the threshold is reached are recorded, which ultimately leads to non-uniform time samples \cite{lazar2004perfect,lazar2003time,lazar2004time}. Time-based sampling hardware, which uses time instances (spikes) only for reconstruction, consumes less power than conventional uniform sampling-based hardware \cite{koscielnik2007designing,rastogi2011integrate,ryu2021time, naaman2023hardware, neuromorphic_computer, camera1}. 
Perfect reconstruction of BL signals from TEM outputs has been widely studied in the literature \cite{lazar2004perfect, lazar2004time,lazar2011video,adam2020sampling,thao2020time,adam2021asynchrony,naaman2021time}. The results on time-encoding machines have also been extended to functions in shift-invariant spaces \cite{gontier2014sampling}, typically by linking time encoding with a non-uniform sampling mechanism \cite{aldroubi2001nonuniform,pnevmatikakis2010spikes}. More recent techniques exploit the signal's underlying structure, allowing accurate sampling and reconstruction at a sub-Nyquist rate \cite{alexandru2019reconstructing,rudresh2020time,naaman2022fri,naamanHW2021}.


Several studies \cite{pnevmatikakis2010spikes,adam2020sampling,alexandru2019reconstructing,naaman2022fri,naaman2022uniqueness} have investigated perfect reconstruction from IF-TEM samples but did not consider the impact of finite quantization. Lazar and T{\'o}th \cite{lazar2004perfect} examined the effects of quantization for BL signals. They showed that the MSE upper bound for time quantization is exactly the same as the MSE upper bound for amplitude quantization when employing non-uniform sampling for the same number of bits. Our analysis, however, charts a different course. We compare the IF-TEM with quantization and the classical ADC that incorporates uniform sampling and quantization. These distinct lines of study can yield different outcomes, primarily due to the different systems being evaluated in relation to the IF-TEM sampler with quantization. 
Moreover, they did not explore the
potential advantages that can be obtained by exploiting the interplay between the signal’s energy, frequency, and maximal amplitude using an IF-TEM sampler with quantization in MSE terms, which is a focal point of our study.


Our contribution is twofold. First, we analyze the quantization step size for the IF-TEM sampler, highlighting its distinctive characteristics when compared to standard sampling techniques. 
Specifically, by leveraging the interplay between the signal's energy, frequency, and maximal amplitude - where an increase in frequency or energy results in a corresponding increase in the signal's maximal amplitude and vice versa -  we demonstrate that as the frequency of the IF-TEM input for BL signals or FRI models increases, the quantization step size decreases.
This analysis sheds light on the unique behavior of the IF-TEM sampler in terms of quantization.
Second, we demonstrate the superior performance of the IF-TEM sampler with quantization, achieving an average of 8 dB improvement compared to uniform classical samplers in terms of MSE for the scenarios considered, encompassing both BL and FRI signal models. 
In particular, we show that this improvement can be achieved by leveraging the interplay between the signal’s characteristics and imposing specific conditions on IF-TEM parameter selection.
Our results show the advantages of the IF-TEM sampler over traditional samplers, leading to reduced bit requirements for both the total number and per sample.

The paper is structured as follows: In Section \ref{sec:encode}, we formulate the problem and provide relevant background information. In Section \ref{sec:REC}, we analyze the quantization strategies for BL signals using classical and IF-TEM sampling techniques. In Section \ref{sec:reconstruction_error}, we present the derivation of recovery error bounds for BL signals resulting from quantization of IF-TEM time instance differences and compare it with classical sampler recovery error resulting from amplitude quantization. Additionally, we derive conditions under which the IF-TEM MSE upper bound is lower than the MSE of a classical sampler. Section \ref{sec:FRI} is devoted to the analysis of quantization strategies for classical and IF-TEM sampling schemes with FRI signals. Finally, we conclude the paper in Section \ref{sec:conc}. 


\begin{figure}[!t]
\begin{center}
\subfigure[]{\includegraphics[width = 2.8in]{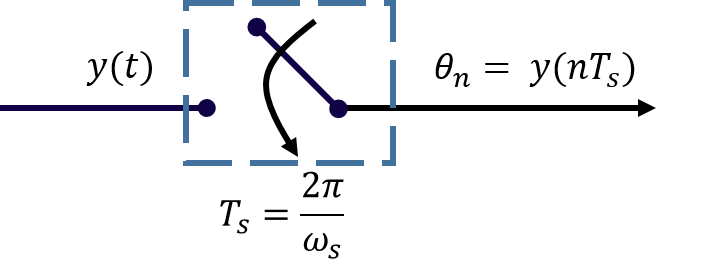}}\vspace{-.1in}
\subfigure[]{\includegraphics[width = 3.2in]{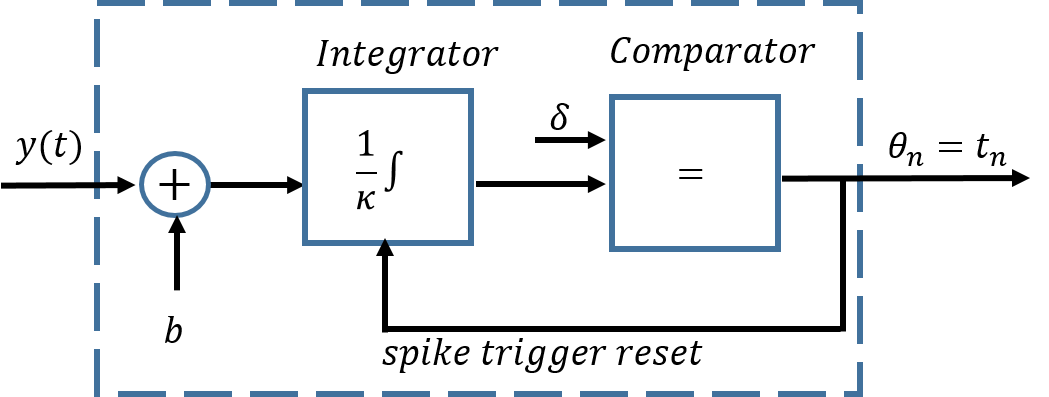}}
\end{center}
\caption{Schematic diagram of (a) classical sampler and (b) IF-TEM sampler. }
    \label{fig:ClassicFRI}
\end{figure}

\section{Preliminaries and Problem formulation}
\label{sec:encode}
We begin with background results for the IF-TEM sampler and discuss signal models. Following that, we present the problem formulation.
\subsection{IF-TEM vs. Classic Sampler}
\label{subsec:IAF}
\begin{figure}[!t]
\begin{center}
\subfigure[]{\includegraphics[width = 3.5in]{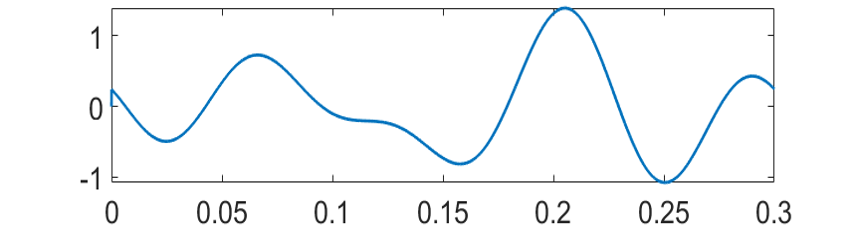}}\vspace{-.1in}
\subfigure[]{\includegraphics[width = 3.5in]{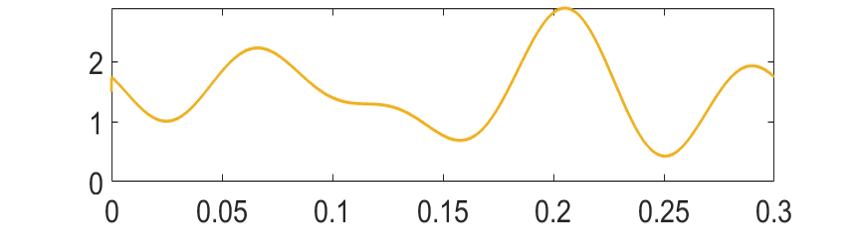}}\vspace{-.1in}
\subfigure[]{\includegraphics[width = 3.5in]{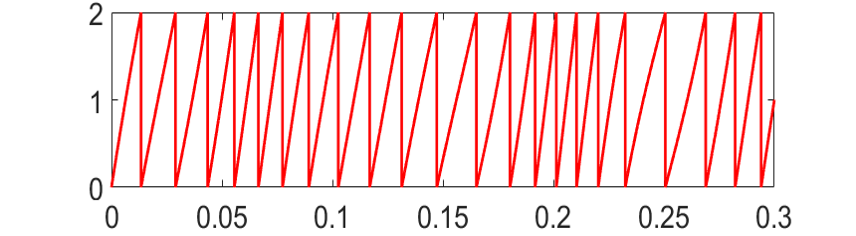}}\vspace{-.1in}
\subfigure[]{\includegraphics[width = 3.5in]{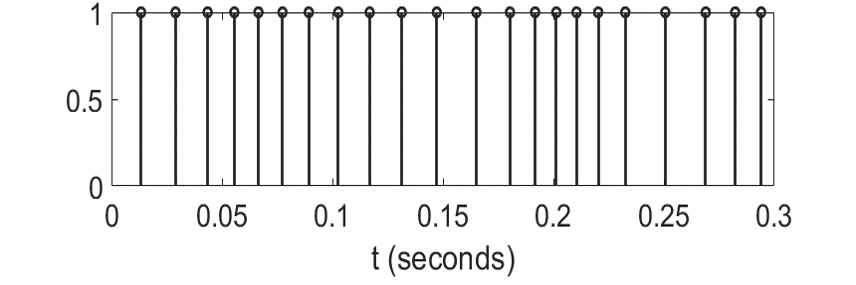}}
\end{center}
\caption{Sampling mechanism of a signal $x(t)$ using an IF-TEM sampler. (a) the signal $x(t)$. (b) the signal with an addition of a bias $b$ such that $x(t)+b$ is a non negative signal. (c) the signal $x(t)+b$ is integrated and scaled, each time the threshold $\delta$ is reached the integrator resets and the time differences between consecutive time instances $T_n = t_n - t_{n-1}$ are recorded. (d) The IF-TEM series of time instances is calculated by summing up the time differences $T_n$ starting from an initial time instant $t_0=0$.  }
\label{fig:TEM_SAMPL}
\end{figure}

Conventional signal sampling involves uniform measurement of the signal's amplitude at specific time intervals. More specifically, given an input signal $x(t)$, the sampling process yields instantaneous samples $x(nT_s)$ with a sampling interval of $T_s$, as shown in Fig. \ref{fig:ClassicFRI}a. In contrast, time encoding machines are a special type of sampling mechanism that encodes the input $x(t)$ by storing time instances instead of amplitudes.
Consider an IF-TEM with bias $b$, scaling $\kappa$, and threshold $\delta$, as shown in Fig. \ref{fig:ClassicFRI}b. The input signal to the IF-TEM, $x(t)$, is real-valued and bounded such that $|x(t)|\leq c<b<\infty$. To time-encode $x(t)$, we add a bias $b$, scale the non-negative signal $x(t)+b$ by $1/\kappa$, and integrate it. The integrator resets when the integral exceeds a threshold $\delta$, and we record the time differences between consecutive firing instants, also called firing intervals (see Fig. \ref{fig:TEM_SAMPL}).  From these firing intervals, we calculate the time instances, which encode the input signal. 

The IF-TEM input $x(t)$ and the time instances $\{t_n\}_{n\in \mathbb{Z}}$ are related as 
\begin{equation}
\int_{t_n}^{t_{n+1}}x(\tau)\, d\tau =
  -b(t_{n+1}-t_n)+\kappa\delta. 
\label{eq:trigger0}
\end{equation}
The measurements $\int_{t_n}^{t_{n+1}}x(\tau)\, d\tau$ are used in the reconstruction of the input signal from the firing instants.
From \eqref{eq:trigger0} and the fact that $|x(t)|$ is bounded by $c$, the time difference $T_n = t_{n+1}-t_n$ is bounded by \cite{lazar2004perfect}
\begin{equation}
 \Delta t_{\min}\triangleq\frac{\kappa\delta}{b+c} \leq T_n \leq \frac{\kappa\delta}{b-c} \triangleq \Delta t_{\max}.
   \label{eq:consecutive_time}
\end{equation}
Due to the bounded nature of temporal time differences, these values are the ones that are quantized and stored in memory.


\subsection{Sampling and Reconstruction of BL Signals}
\label{subsec:rec_IF}
Signal reconstruction from IF-TEM outputs has been established when the input is a $c$-bounded, $2\Omega$ BL signal in $\mathbf{L}^2(\mathbb{R})$ \cite{lazar2004perfect,lazar2004time,adam2020sampling,naaman2021time}.
\begin{definition}
A signal $x(t)$ is said to be $c$-bounded and $2\Omega$ BL signal if 
$|x(t)|\leq c$, where $c\in\mathbb{R}$, and its Fourier transform is zero outside the closed interval $[ -\Omega,\Omega]$, where $\Omega$ is in radians per second. 
\end{definition}
The Shannon-Nyquist theorem, which we refer to as the classical approach, states that a $2\Omega$ BL signal $x(t)$ can be perfectly recovered from its uniform samples $x(nT_s)$, if the sampling rate is at least the Nyquist rate $\frac{\Omega}{\pi}$ Hz \cite{nyquist1928certain}.
\begin{definition}
A signal $x(t)\in \mathbf{L}^2(\mathbb{R})$ is said to have finite energy $E\in\mathbb{R}$ if  
\[
E = \int_{-\infty}^{\infty} |x(t)|^2dt <\infty.
\]
\end{definition}
An effective way to express the energy of a BL finite-energy function is by its Parseval's relation, which produces
\begin{equation}
    E = \int_{-\infty}^{\infty} |x(t)|^2dt = \frac{1}{2\pi}\int_{-\Omega}^{\Omega}|X(j\omega)|^2d\omega,
\end{equation}
where $X(j\omega)$ is the continuous-time Fourier transform of the signal $x(t)$.
In general, the bandwidth $2\Omega$ and the amplitude upper-bound $c$ are independent. Here, we consider BL signals with maximal energy $E$; in this case, as given in \cite{papoulis1967limits}, the relation between $\Omega$ and $c$ is 
\begin{equation}
    c = \sqrt{\frac{E\Omega}{\pi}}.
    \label{eq:c_omega_connection}
\end{equation}

We adopt the IF-TEM sampling mechanism with a zero refractory period, following the approach proposed in \cite{lazar2004time}. By using an iterative approach, the authors in \cite{lazar2004time,pnevmatikakis2010spikes} showed that BL signals can be perfectly recovered using an IF-TEM with parameters $\{b,\kappa,\delta\}$ if $b>c$ and 
\begin{equation}
    \Delta t_{\max} = \frac{\kappa\delta}{b-c} < \frac{\pi}{\Omega}.
    \label{eq:density}
\end{equation}
The bound in \eqref{eq:density} requires a bandwidth that is inversely proportional to the time difference between the firing instants, i.e., the BL input signal can be recovered if the overall firing rate of the IF-TEM
is higher than the Nyquist rate.
Using the IF-TEM sampler, the signal is reconstructed similarly to a BL signal with non-uniformly spaced amplitude samples. 
The reconstruction operator $\mathcal{R}$ is defined as:
\begin{equation}
    \mathcal{R}(x(t)) = \sum_{i=1}^{\infty} \left(\int_{t_i}^{t_{i+1}} x(u)\mathrm{d}u\right)\sinc_\Omega(t-s_i),
    \label{eq:6}
\end{equation}
where $s_i=\frac{t_i+t_{i+1}}{2}$ and 
\begin{equation}
    \sinc_\Omega(t)=
    \begin{cases}
    \frac{\sin(\Omega t)}{\pi t}, & \text{if $t\neq 0$}.\\
    1, & \text{otherwise}.
     \end{cases}
     \label{eq:sinc}
\end{equation}
The values of $\int_{t_i}^{t_{i+1}} x(\tau)d\tau$ $ i\in\mathbb{Z}$ are evaluated from the IF-TEM output sequence $\{t_i\}_{i\in\mathbb{Z}}$ using \eqref{eq:trigger0}.
Assuming \eqref{eq:density} holds, the $c$-bounded $2\Omega$ BL signal $x(t)$ can be perfectly recovered using the following iterative algorithm, as showed in \cite{lazar2004perfect}:
\begin{equation}
x_{l+1}(t) = x_l(t) + \mathcal{R}\left(x(t)-x_l(t)\right),
\end{equation}
where $l\in\mathbb{N}$ and $x_0(t) = \mathcal{R}(x(t))$.
It is crucial to note that the signal $x(t)$ is known in analog form rather than digital form. Therefore, the subtraction $x-x_l$ can be generated using analog hardware and then passed through $\mathcal{R}$. In this case, $\lim_{l\to\infty} x_l(t) = x(t)$, and $\lVert x-x_l \rVert \leq r^{l+1}\lVert x \rVert$, where $r =\frac{\kappa\delta}{b-c} \frac{\Omega}{\pi}<1$.
Next, we discuss sampling and recovery for FRI signals.

\subsection{Sampling and Reconstruction of FRI Signals}
\begin{figure}[t!]
		\centering
		\includegraphics[width= 3.5
		in]{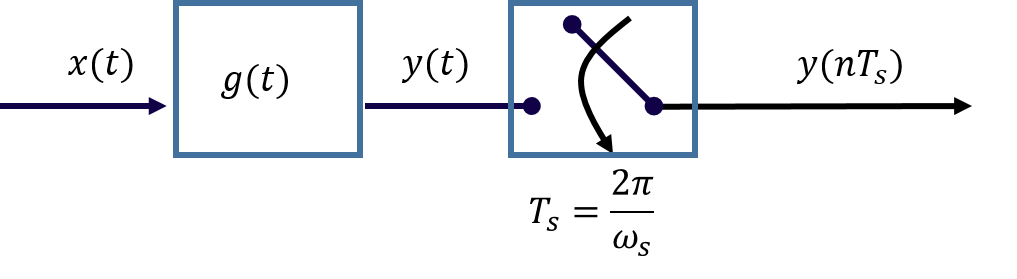}
	    \vspace*{-0.7cm}
		\caption{A kernel-based FRI sampling framework: An FRI signal $x(t)$ is first filtered by a sampling kernel $g(t)$ and then instantaneous uniform samples are measured at a sub-Nyquist rate. Parameters of the FRI signal are estimated from the sub-Nyquist samples.}
		\label{fig:FRIgeneric}
	\end{figure}

Sampling and recovery of FRI signals, which have a limited number of degrees of freedom, are of great interest in applications such as radar and ultrasound \cite{eldar2015sampling,vetterli2002sampling,tur2011innovation}. Common FRI signal models involve a linear combination of delayed copies of a known pulse. A typical framework for sampling and reconstruction of such signals includes a custom sampling kernel, an ADC, and a parameter estimation block. The purpose of the sampling kernels is to disperse the FRI signal information so that the parameter estimation block can estimate time delays and amplitudes using a low sampling rate and a finite number of samples (see Fig. \ref{fig:FRIgeneric}).
In \cite{naaman2022fri}, the recovery of both periodic and non-periodic FRI signals using the IF-TEM sampler is demonstrated. Here, we focus on sampling and reconstruction of $T$-periodic FRI signals.

A signal $x(t)$ is said to be a $T$-periodic FRI signal if
\begin{equation}
    x(t) = \sum_{p\in\mathbb{Z}}\sum_{\ell=1}^L a_{\ell} h(t-\tau_{\ell}-p T),
    \label{eq:fri}
\end{equation}
where the FRI parameters $\{(a_{\ell},\tau_{\ell})|\tau_{\ell} \in (0, T],a_{\ell}\in\mathbb{R} \}_{\ell=1}^L$ are the unknown amplitudes and delays. The number of FRI pulses $L$ and the pulse shape $h(t)\in\mathbf{L}^2(\mathbb{R})$ are known. 

Since $x(t)$ is $T$-periodic, it has the following Fourier series representation 

\begin{equation}
    x(t) = \sum_{k\in\mathbb{Z}}X[k]e^{j k \omega_0 t},
\end{equation}
where 
\begin{equation}
 X[k] = \frac{1}{T}H(k\omega_0)\sum_{\ell=1}^{L}a_{\ell}e^{-jk \omega_0 \tau_{\ell}},
 \label{eq:x_initial,eq:x_fourier}
\end{equation}
and $\omega_0 = \frac{2\pi}{T}$. Here $H(\omega)$ is the continuous-time Fourier transform of $h(t)$.
The rate of innovation of $x(t)$, that is, the degrees of freedom per unit time interval, is $\frac{2L}{T}$.

The parameters $\{a_{\ell}, \tau_{\ell}\}_{\ell=1}^L$ can be uniquely computed from a minimum of $2L$ and $2L+2$ Fourier series coefficients (FSCs), for the classical and IF-TEM setup respectively, by using spectral analysis methods, such as the annihilating filter (AF) \cite{vetterli2002sampling,eldar2015sampling,rudresh2020time,naaman2022fri}. Hence, $2L$ FSCs $X[k]$ can uniquely determine the FRI signal $x(t)$ \cite{eldar2015sampling}. Thus, the FRI signal reconstruction problem is reduced to uniquely determining the desired number of FSCs from the signal measurements.
Reconstruction from IF-TEM outputs has been considered for cases where the FRI input is $c$-bounded and is guaranteed if the IF-TEM parameters satisfy \cite{naaman2022fri}
\begin{equation}
    \frac{1}{\Delta t_{\max}}\geq \frac{2L+2}{T}.
    \label{eq:FRI_rec}
\end{equation}  
This condition is similar to the classical FRI method where perfect recovery is achieved when sampling at the rate of innovation.

\begin{figure}
    \centering
    \includegraphics[width=0.5\textwidth]{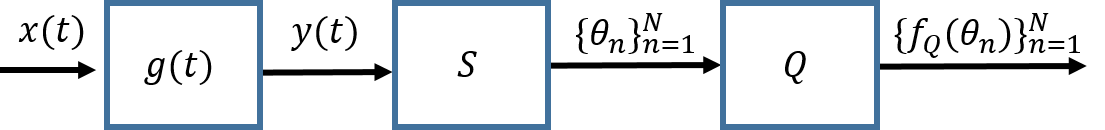}
    \caption{Generalized sampling with quantization system mode: Filtered continuous-time signal $y(t)=(x*g)(t)$ is sampled by a sampler $S$ that results in a discrete representation  $\{\theta_n\}_{n\in I}$, where $I$ is a countable set. The representation is quantized by a quantizer $Q$, resulting in $\{f_Q(\theta_n)\}_{n\in I}$.}
    \label{fig:IAF}
\end{figure}
\subsection{Problem Formulation}
Previous studies have predominantly concentrated on perfect sampling and reconstruction using IF-TEM, disregarding the impact of quantization. The authors in \cite{lazar2004perfect} compared the reconstruction errors resulting from quantizing in the time and amplitude domains with non-uniform sampling. 
They showed that, with non-uniform sampling employed for amplitude quantization, the MSE upper bound for time quantization exactly matches that of amplitude quantization for the same number of bits.
However, these investigations did not include a comparison between the IF-TEM sampler with quantization and the classical uniform sampler, nor did they explore the relationship between signal energy, frequency, and maximal amplitude. Additionally, the consideration of FRI signals was not part of their analyses.
In contrast, our contribution addresses these gaps by conducting a comparison between the IF-TEM sampler with quantization and the classical uniform sampler. Additionally, we examine the interplay between signal energy, frequency, and maximal amplitude to shed light on its impact on the quantization step size for both the IF-TEM and the classical sampler.


Our objective is to capitalize on this relationship and analyze 
the performance in terms of MSE. Specifically, we aim to 
evaluate and compare the signal recovery error for both the IF-TEM and classical setups when reconstructing a signal $x(t)$, which can be either an FRI or a BL signal, from its quantized samples.
For BL signals, we consider a $2\Omega$ BL and $c$-bounded signal in $\mathbf{L}^2(\mathbb{R})$ with a maximum energy of $E<\infty$. For FRI signals, we focus on a $T$-periodic FRI signal as defined in \eqref{eq:fri}, where our goal is to retrieve the FRI parameters $\{a_{\ell},\tau_{\ell}\}_{\ell=1}^L$ from quantized IF-TEM samples, with FSCs calculated from quantized samples.

A generalized sampling with a quantization scheme is illustrated in Fig. \ref{fig:IAF}. The input signal $x(t)$ is passed through a sampling kernel $g(t)$, resulting in the signal $y(t)$. From $y(t)$, we compute a set of measurements, $\{\theta_n\}_{n\in I}$, where $I$ is a countable set. The \emph{sampler} is denoted by \emph{S}, and the \emph{quantizer} is denoted by \emph{Q}. The sampler can be an instantaneous classical sampler, as shown in Fig. \ref{fig:ClassicFRI}(a), with $\theta_n = y(nT_s)$, or an IF-TEM, as depicted in Fig. \ref{fig:ClassicFRI}(b), with $\theta_n = t_n$.

For BL input signal, the sampling kernel is a low-pass filter or can be removed. For FRI input signal, in both IF-TEM and classical settings, the sampling kernel $g(t)$ is designed such that $2L+2$ FSCs of $x(t)$ are computed from $\{\theta_n\}_{n\in I}$, where $|I|\geq 2L+2$. A sum-of-sincs (SoS) filter can be used to determine the FSCs from the measurements \cite{naaman2022fri,tur2011innovation}.

After the sampler computes the measurements $\theta_n$ of the signal $x(t)$, the samples are quantized. In the classical ADC, the instantaneous amplitude samples $y(nT_s)$ are quantized using a uniform quantizer. In the quantized IF-TEM sampler, the differences of time-encodings, denoted $T_n = t_{n+1}-t_n$ are discretized using a uniform quantization\footnote{As explained in Section \ref{sec:REC}, we also compared the IF-TEM results to the classical sampling scheme where the instantaneous samples $y((n+1)T_s)-y(nT_s)$ are quantized.}. We shall refer to these quantized time differences as $\hat{T}_n$.



In the following sections, we begin by examining the quantization strategies employed in classical and IF-TEM sampling schemes for BL signals. The methodology on how to choose the appropriate quantization step size is discussed. Subsequently, we perform a comparative analysis of the classical and IF-TEM reconstructions by deriving an upper bound on the MSE of the recovered BL signals from the quantized measurements $f_Q(\theta_n)$. Furthermore, we identify the specific parameter settings for the IF-TEM sampler that result in superior MSE performance compared to the classical sampler while maintaining a fixed and finite quantization resolution. We then demonstrate the superior MSE performance compared to the classical sampler, extending our findings to FRI signals.
\section{Quantization Analysis}
\label{sec:REC}
In this section, we study quantization for classical and IF-TEM sampling schemes for BL signals.
\subsection{Classical and IF-TEM Samplers Quantization Step}
We first demonstrate that as in IF-TEM the energy $E$ or the frequency of the BL signal grows, the dynamic range $\Delta t_{{max}}-\Delta t_{{min}}=\frac{\kappa\delta}{b-c}-\frac{\kappa\delta}{b+c}$ of each time sample decreases. 
In particular, we show that, in contrast to the classical method, for a specific selection of IF-TEM parameters and fixing the ratio between $b$ and $c$, increasing the frequency or energy of the signal increases the quantizer’s resolution \cite{naaman2021time}.

In either the traditional or IF-TEM sampling system, the sampled signal is quantized by an identical uniform scalar quantizer with a resolution of $\log_2 K$ bits, meaning that each quantizer can produce $K$ distinct output values \cite{gray1998quantization}. We begin by discussing quantization of a BL signal within the classical framework. Let $x(t)$ be a BL signal with finite energy $E$ \cite{papoulis1967limits},
\begin{align}
    |x(t)| \leq c \triangleq \sqrt{\frac{E\Omega}{\pi}}.
    \label{eq:cenergy}
\end{align}
Since  $x(t)$ is $c$-bounded, the dynamic range of the instantaneous samples $x(nT_s)$ lie within $[-c, c]$. Consider a $K$ level uniform quantizer with $N$ bits, i.e., $K=2^N$. The classical sampler quantization step size is given by
\begin{align}
    \Delta_{\text{classic}} = \frac{2c}{K}.
    \label{eq:classic_stepsize}
\end{align}

It is worth noting that when quantizing the amplitude differences, akin to quantizing the time differences in IF-TEM, the range of the instantaneous samples $x((n+1)T_s) - x(nT_s)$ lies within the interval $[-2c, 2c]$. In this scenario, the quantization step size for the amplitude differences in the classical sampler can be expressed as:
\begin{equation}
\Delta_{\text{classic diff}} = \frac{4c}{K} = 2\Delta_{\text{classic}}.
\label{eq:classic_stepsize_diff}
\end{equation}
Consequently, by utilizing the same number of bits $N=\log_2 K$, the quantization step resolution for the amplitude differences method is inferior to that of the classical method, where only the amplitudes themselves are quantized.
In light of this, we proceed to compare the quantization performance of the IF-TEM approach with the quantization employed in the classical method.

In the IF-TEM sampler, the time-differences $T_n$ are quantized. Using \eqref{eq:consecutive_time}, the dynamic range of $T_n$ is $[\frac{\kappa\delta}{b+c},\frac{\kappa\delta}{b-c}]$. Hence, for a $K$-level uniform quantizer, the IF-TEM quantization step-size is given by
\begin{equation}
    \Delta_{\text{IF-TEM}} = \frac{\frac{\kappa\delta}{b-c}-\frac{\kappa\delta}{b+c}}{K} = \frac{\kappa\delta}{(b+c)(b-c)} \frac{2c}{K}.
    \label{eq:iftem_stepsize}
\end{equation}

We note that the integrator constant $\kappa$ is a parameter of the integrator circuit, which is usually fixed \cite{adam2020sampling,naaman2023hardware}. In practice, the threshold $\delta$, which is a parameter of the comparator, and the bias $b$, are easier to control. 
Thus, to have sufficient number of samples for recovery \eqref{eq:density}, one can increase the bias $b$ or decrease the threshold $\delta$.
Both $b$ and $\delta$ are generated by a DC voltage source, so large values of bias and threshold necessitate a high power input. Therefore, to minimize power requirements, the bias $b$ and $\delta$ should be as small as possible. Nevertheless, for fixed values of $b$ and $\kappa$, low $\delta$ results in a large firing rate, above the minimum desirable firing rate. Thus, a possible way to strike a balance is to select fixed values of $\delta$ and $\kappa$, and to choose the bias $b>c$ such that $b= \alpha c$, where $\alpha>1$ and $b-c = \epsilon>0$. Consequently, to analyze the relation between the signals frequency $\Omega$ and $\Delta_{\text{IF-TEM}}$, fixed values of $\delta$ and $\kappa$ are assumed, while $b$ may vary.

We next show that by increasing $\Omega$, the quantization step size $\Delta_{\text{IF-TEM}}$ decreases.
We summarize this result in the following theorem.  
\begin{theorem}
\label{theorem:IF_Quant_BL}
Consider an IF-TEM sampler, succeeded by a $K$-level uniform quantizer, where $K=2^N$ and $N$ denotes the number of bits. For $2\Omega$-BL signals with maximal energy $E$, given a fixed $\alpha > 1$, let the IF-TEM bias be represented by $b$, such that $b = \alpha c$, where $c$ is defined in \eqref{eq:cenergy}. Under Nyquist-like constraint \eqref{eq:density}, the quantization step $\Delta_{\text{IF-TEM}}$ decreases as the energy $E$ or the frequency $\Omega$ increases.
\end{theorem}

\begin{proof}
Fix $\kappa$ and $\delta$. The bias is chosen such that $b = \alpha c$ with fixed $\alpha >1$. Substituting $b$ into \eqref{eq:iftem_stepsize}, we have
\begin{equation}
    \Delta_{\text{IF-TEM}} = \frac{\kappa\delta}{(\alpha+1)(\alpha-1)} \frac{2}{cK}.
    \label{eq:iftem_stepsize22}
\end{equation}
Using condition \eqref{eq:cenergy} and \eqref{eq:density} results in
\begin{equation}
     \Delta_{\text{IF-TEM}} = \frac{\kappa\delta}{(\alpha+1)(\alpha-1)} \frac{2}{K}. \sqrt{\frac{\pi}{E\Omega}}.
\end{equation}
Thus, with an increasing signal's energy $E$ or frequency $\Omega$, the IF-TEM quantization step size decreases.
\end{proof}

These findings highlight a significant difference from the classical method: Increasing the signal's frequency or energy, which correlates with the maximal amplitude of the class $c$, leads to an improved resolution of the quantizer. The connection between $b$ and $c$ is pivotal, particularly the relationship $b = \alpha c$ for a constant $\alpha > 1$. This relationship is instrumental in deducing that $\Delta_{\text{IF-TEM}}$ decreases as the signal amplitude increases. Note that an increase in $c$ does not necessarily decrease the quantization step size $\Delta_{\text{IF-TEM}}$.

\begin{figure}[h!]
	\centering
	\includegraphics[width=0.45\textwidth]{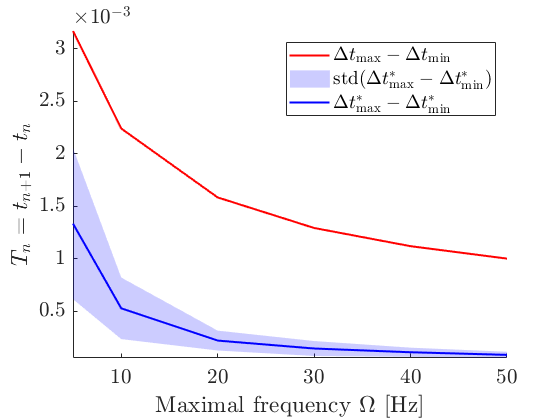}
	\caption{Time instances differences in IF-TEM with BL signals as a function of the frequency band. In red: the difference  $\Delta t_{\max} - \Delta t_{\min}$, as defined in \eqref{eq:consecutive_time}. In blue: the solid line shows the average values of $\Delta t_{\max}^{*} - \Delta t_{\min}^{*}$, which are the real difference. See the range in Table~\ref{Tab:Tcr2}.}
	\label{fig:MSE12bits}
\end{figure}
\begin{table}[h!]
    \centering
    \caption{Time instants range in IF-TEM with BL signals.}
    \begin{tabular}{ccccccc}
        \hline
         Frequency [HZ] &$5$&$10$&$30$&$50$  \\
         \hline
         $\Delta t_{\max} - \Delta t_{\min}$[$e{-04}$]&$32$&$22$&$13$&$10$\\
         \hline
         $\Delta t_{\max}^{*} - \Delta t_{\min}^{*}$[$e{-04}$]&$13$&$5.2$&$1.4$&$0.85$\\
         \hline
    \end{tabular}
    \label{Tab:Tcr2}
\end{table}
To underscore the significance of this relationship, consider an example with a non-fixed $\alpha > 1$, where $b = \alpha c$. Given a fixed $K$ (corresponding to a fixed number of bits per sample, using uniform quantizer) and choosing $\kappa = \delta = b = 1$:
In the first case: For $c= \frac{1}{4}$, this corresponds to $\alpha = 4$, resulting in $\Delta_{\text{IF-TEM}} = \frac{8}{15K}$.
In the second case: For $c = \frac{1}{2} > \frac{1}{4}$ (indicating an increase in $c$), this relates to $\alpha = 2$, yielding $\Delta_{\text{IF-TEM}} = \frac{4}{3K}$.
Despite the increase in $c$, we observe an increase in the quantization step size. This result is attributed to our choice of a non-fixed $\alpha$.
The relation between the amplitude $c$ and the bias $b$ needs to obey the Nyquist condition given in \eqref{eq:density}. Moving forward, we assume that the relationship between $b$ and $c$ is defined as $b = \alpha c$, with a constant $\alpha > 1$.

Comparing \eqref{eq:classic_stepsize}, \eqref{eq:classic_stepsize_diff}, and \eqref{eq:iftem_stepsize22}, we observe that the quantization step size in classical sampling increases with the amplitude of the BL signal. However, in IF-TEM, the step size decreases with amplitude. 
Furthermore, when increasing $\Omega>0$, the time instances become closer, which causes smaller values of $T_n$s and $\Delta t_{\min},\Delta t_{\max}$ become closer (see Fig.~\ref{fig:MSE12bits} and Table~\ref{Tab:Tcr2}).
Thus, as shown in Section \ref{sec:reconstruction_error}, the quantization error can be reduced, and the IF-TEM results in a lower quantization error than the classical scheme for the same amount of bits.

As a final remark, for fixed $K$, increasing the signal's energy $E$ or frequency $\Omega$ for the BL signal increases the number of samples $N$ in both IF-TEM and classical methods.  Thus, the total number of bits increases in both methods. For comparison, we set the number of samples to be equal for both classical and IF-TEM. 

\subsection{Evaluation results}
\label{sec:sim}

We now present empirical evidence to support our conclusion that, for a fixed number of quantization bits, the quantization step size can be reduced when either the bandwidth or frequency of the BL signal grow.


Consider a $2\Omega$-BL input signal $x(t)$ which is bounded in amplitude, i.e., $|x(t)|\leq c$, for $c=\sqrt{(E\Omega)/\pi}$ with $E=1$ and $\Omega$ varying from $5-50$ Hz. 
We investigate the recovery after quantization for the IF-TEM sampler.
The input signal is given by 
\begin{equation}
    x(t) = \sum_{n=-i}^i a[n] \sinc\left(\frac{t-nT_s}{T_s}\right),
\end{equation}
where $i=3$, $T_s = \frac{1}{2\Omega}$, and $a[n]$ are 100 uniform i.i.d sets within the range $[-1,1]$. Subsequently, the signal is normalized to have an energy of $E=1$.
The IF-TEM parameters are selected as follows; we use fixed values of $\delta=0.075$ and $\kappa=0.4$. To have a sufficient number of samples for recovery, the bias is selected such that $b=4c$, i.e, $\alpha=4$, and satisfies the condition $|x(t)|\leq c$. The uniform quantizer used has $N=12$ bits.
We demonstrate that the time instances differences and their range, $\Delta t_{\min} - \Delta t_{\max}$, decreases as the frequency of the signal increases.
The results are shown in Fig.~\ref{fig:MSE12bits} and Table~\ref{Tab:Tcr2}.

Since the IF-TEM parameters, $\kappa$ and $\delta$ are chosen to be constant, and $b= 4c$, from \eqref{eq:consecutive_time} and \eqref{eq:c_omega_connection}, it is evident that only the maximal amplitude $c=\sqrt{(E\Omega)/\pi}$, or the frequency $\Omega$, affects the interval $\Delta t_{\min} - \Delta t_{\max}$ (see Fig.~\ref{fig:MSE12bits}).
In order to examine the frequency influence not only on the boundaries but also on the signal itself, based on the fact that $|x(t)|\leq c$, we define tighter bounds on the difference between the time instances as follows:
\begin{equation}
    \Delta t_{\max}^{*} = \frac{\kappa\delta}{b-\max(x(t))}\leq\frac{\kappa\delta}{b-c}=\Delta t_{\max},
\end{equation}
and
\begin{equation}
    \Delta t_{\min}^{*} = \frac{\kappa\delta}{b+\max(x(t))}\geq\frac{\kappa\delta}{b+c}=\Delta t_{\min}.
\end{equation}
Note that $\Delta t_{\max}^{*}-\Delta t_{\min}^{*}\leq\Delta t_{\max} - \Delta t_{\min}$. Thus, as the BL signals frequency is higher, one can increase the resolution of the quantizer using the same number of $12$ bits.


\section{Reconstruction error with quantization}
\label{sec:reconstruction_error}
Next, utilizing Theorem \ref{theorem:IF_Quant_BL}, which shows that the quantization error $\Delta_{\text{IF-TEM}}$ decreases with increasing amplitude $c$ or frequency $\Omega$ of the BL signal, we demonstrate that the IF-TEM method achieves a lower MSE compared to the conventional ADC approach.
\subsection{IF-TEM Quantization Noise}
In this section, we introduce an upper bound for the error resulting from time sequence quantization. Lazar and T{\'o}th \cite{lazar2004perfect} established an upper bound for the reconstruction error associated with the asynchronous sigma-delta modulator (ASDM) sampler. They demonstrated that the MSE upper bound for time quantization is the same as that of amplitude quantization under conditions of non-uniform sampling for both samplers with a fixed number of bits. Our work, however, offers a distinct perspective. We evaluate the MSE of the IF-TEM with quantization, and compare it with the conventional ADC that employs uniform sampling and quantization. Given the distinctions in our methodological approaches, it is anticipated that our results may differ, particularly when considering the specific characteristics of the IF-TEM sampler with quantization we introduce here. Note that Lazar and T{\'o}th \cite{lazar2004perfect} did not explore the relationship between a signal’s energy, frequency, and maximal amplitude in the context of the time-based sampler MSE. Yet, even if this aspect had been investigated, their conclusions, anchored in non-uniform sampling for both ASDM and traditional samplers, would remain the same. In contrast, we show that the quantization step size of the IF-TEM sampler can be decreased when the maximal frequency of a BL signal is increased or the number of pulses of an FRI signal is increased. Consequently, under specific parameter configurations, the IF-TEM sampler exhibits lower MSE bound compared to a classical ADC with equivalent bit depth.

In the following, we introduce an upper bound for the IF-TEM MSE, as described in Theorem \ref{thm:Ebound} building upon the relationships between the signal's energy, frequency, and maximal amplitude \eqref{eq:c_omega_connection}, as well as the connection between the signal's maximal amplitude $c$ and the IF-TEM bias $b$, that is, $b=\alpha c$ with a constant $\alpha > 1$.  This theorem also draws from Theorem \ref{theorem:IF_Quant_BL}, which demonstrates that the quantization step $\Delta_{\text{IF-TEM}}$ decreases as the maximal signal amplitude $c$, or frequency $\Omega$, increases.
Our analysis reveals that the IF-TEM sampler can achieve superior results compared to the conventional sampler in terms of MSE, using the same number of samples and bits.

Consider the scenario where the sequence of time instances $\{t_i\}_{i\in\mathbb{Z}}$ is subject to measurement with limited precision. Denote the recovered values by $\{\hat{t}_i\}_{i\in\mathbb{Z}}$. We define $T_i = t_{i+1}-t_i$ and $\hat{T}_i = \hat{t}_{i+1}-\hat{t}_i$ for all $i\in\mathbb{Z}$.
Assume that $\hat{\mathcal{R}}$ is a reconstruction operator defined by (see \eqref{eq:6}):

\begin{equation}
    \hat{\mathcal{R}}(x(t)) = \sum_{i=1}^{\infty} \left(\int_{\hat{{t_i}}}^{\hat{t}_{i+1}} x(u)\mathrm{d}u\right)\sinc_\Omega(t-\hat{s}_i),
\end{equation}
where $\hat{s}_i = \dfrac{\hat{t}_{i+1}+\hat{t}_i}{2}$ and $\sinc_\Omega(t)$ is defined in \eqref{eq:sinc}.
In this case, it is was shown in \cite{lazar2004perfect} that the following expressions holds:
\begin{equation}
    x(t) = \sum_{k\in \mathbb{N}} (I-\hat{\mathcal{R}})^k \hat{\mathcal{R}}x(t),
    \label{eq:x}
\end{equation}
and
 \begin{equation}
      \hat{x}(t) = \sum_{i\in \mathbb{N}}(I-\hat{\mathcal{R}})^i \sum_{l\in Z}[\kappa\delta-b\hat{T}_l]\sinc_\Omega(t-\hat{s}_l)).  
      \label{eq:xhat}
 \end{equation}
The reconstruction algorithm yields a consistent error signal represented by $e(t) = x(t) - \hat{x}(t)$ denote the signal error. Using \eqref{eq:x} and \eqref{eq:xhat}, we can express the error as a summation:
\begin{equation}
e(t) = \sum_{k\in Z}(I-\hat{\mathcal{R}})^k \sum_{l\in Z}\epsilon_l \sinc_\Omega(t-\hat{s}_l),
\label{eq:e}
\end{equation}
where 
\begin{equation}
\epsilon_l = [\kappa\delta - b\hat{T}_l] - \int_{\hat{t}_l}^{\hat{t}_{l+1}} x(u)\mathrm{d}u.
\label{eq:eps}
\end{equation}

When there is no quantization error, we have $\hat{t}_l = {t}_{l}$. Substituting this into \eqref{eq:eps}, and using \eqref{eq:trigger0} leads to $\epsilon_l = 0$ and $e(t)=0$. However, with quantization, $\hat{t}_l$ may not necessarily be equal to $t_l$ and $\hat{T}_l$ may not necessarily be equal to $T_l$, rendering \eqref{eq:trigger0} inapplicable, and thus $\epsilon_l$ may not be zero.
Considering the definition of \eqref{eq:eps}, we observe that since $x(t)$ is BL, $\int_{\hat{t}_l}^{\hat{t}_{l+1}} x(u)\mathrm{d}u\in \ell^2$, while $[\kappa\delta - b\hat{T}_l]$ may not belong to $\ell^2$ due to the quantization of the time differences. In the absence of quantization, $[\kappa\delta - b{T}_l]  = \int_{{t}_l}^{{t}_{l+1}} x(u)\mathrm{d}u\in \ell^2$. Consequently, $e(t)$ as defined in \eqref{eq:e} may not necessarily be in $L^2(\mathbb{R})$ since it is a series of shifted sinc functions with coefficients that may not be in $\ell^2$. 
Hence, computing the MSE by employing an $L_2$ norm on \eqref{eq:e} is not a straightforward task. Instead, we follow \cite{lazar2004perfect} and define a squared error measure $\varepsilon^2$ of the quantization process as follows
\begin{equation}
    \varepsilon^2 = \lim_{n\rightarrow\infty} \frac{1}{2n\Delta t_{\min}} {||e \mathds{1}_{[-n\Delta t_{\min},n\Delta t_{\min}]}||}^2,
    \label{eq:epsilon_square}
\end{equation}
where $\mathds{1}_{[-n\Delta t_{\min},n\Delta t_{\min}]}$ is an indicator function, $\Delta t_{\min}$ is the minimum width between any two consecutive time instants, $T_k$ is the time difference between the $k$-th and $(k+1)$-th time instants, and
\begin{equation}
    ||e\mathds{1}_{[-n\Delta t_{\min},n\Delta t_{\min}]}||^2 = \int_{\mathbf{R}} e^2(t) \mathds{1}_{[-n\Delta t_{\min},n\Delta t_{\min}]}(t)dt.
\end{equation}

If the error in quantizing the time differences is modeled as a random process, $\epsilon_l$  becomes a random quantity. Consequently, $\varepsilon^2$ and $e(t)$ are also random. Therefore, we focus on the expectation, $\EX[\varepsilon^2]$, which we will refer to as the MSE. In the following theorem, we introduce an upper bound for the MSE of the IF-TEM $\EX[\varepsilon^2]$ when the quantization error on the recorded time differences is modeled as a uniformly distributed i.i.d. sequence.

\begin{theorem}
\label{thm:Ebound}
Let $x(t)$, $t\in\mathbb{R}$ be a signal BL to $[-\Omega,\Omega]$. Consider an IF-TEM sampler followed by a $K$-level uniform quantizer with $N = \log_2 K$ bits. Let $E$ be the maximal energy of $x(t)$, and let the relationship between $E$, $\Omega
$ and $c$ be given by $c = \sqrt{{E\Omega}/{\pi}}$. 
Let $b = \alpha c$ for any fixed $\alpha>1$, where $b$ represents the IF-TEM bias.
Consider a sequence of uniform i.i.d random variables $d_k = \hat{T}_k -T_k$ on $[-\frac{\Delta_{\text{IF-TEM}}}{2}, \frac{\Delta_{\text{IF-TEM}}}{2}]$ as the quantization error, where $\Delta_{\text{IF-TEM}}$ defined in \eqref{eq:iftem_stepsize22}.
Then, the MSE is upper bounded by
\begin{equation}
\EX[\varepsilon^2] < \dfrac{R}{(1-R)^2} \left(\dfrac{\alpha+1}{\alpha-1}\right)\left(\dfrac{E\Omega}{3\pi}\right) \dfrac{1}{2^{2N}},
\label{eq:UB_bits}
\end{equation}
where $R = \left(\dfrac{\kappa\delta}{\alpha-1}\right) \sqrt{\frac{\Omega}{E\pi}}<1$.
\end{theorem}
\begin{proof}
Let $x(t)$, where $t \in \mathbb{R}$, be a BL signal to $[-\Omega,\Omega]$. The maximal amplitude of this signal is denoted by $c$. Consider an IF-TEM sampler followed by a $K$-level uniform quantizer using $N = \log_2 K$ bits. Initially, we focus on proving Lemma \ref{thm0}, for which the proof can be found on Appendix \ref{app:prop1}. In this scenario, the relation $b=\alpha c$ with a fixed $\alpha>1$, and the relationship between $E$, $\Omega$, and $c$ which is given by $c = \sqrt{{E\Omega}/{\pi}}$, is not considered. The quantization step size in this case follows from \eqref{eq:iftem_stepsize}. The parameters of the IF-TEM, $\{\kappa, b>c,\delta\}$, are solely selected to adhere to the Nyquist criterion \eqref{eq:density}, that is, $r = \frac{\kappa\delta\Omega}{\pi (b-c)}<1$.

\begin{lemma}
\label{thm0}
Suppose $d_k = \hat{T}_k -T_k \sim \mathrm{Unif}(-\frac{\Delta_{\text{IF-TEM}}}{2}, \frac{\Delta_{\text{IF-TEM}}}{2})$ is the quantization error, where $\Delta_{\text{IF-TEM}}$ defined in \eqref{eq:iftem_stepsize}. The MSE $ \EX[\varepsilon^2]$, which $\varepsilon^2$ given in \eqref{eq:epsilon_square}, can be upper bounded as
\begin{equation}
    \EX[\varepsilon^2] < \frac{b+c}{T\kappa\delta} \left(\frac{b+c}{1-r}\right)^2 \frac{\Delta_{\text{IF-TEM}}^2}{12}.
    \label{eq:UB1}
\end{equation}
\end{lemma}

When considering a fixed ratio between $b$ and $c$ given by $\alpha=\frac{b}{c}>1$, the IF-TEM quantization step size from Theorem \ref{theorem:IF_Quant_BL} is valid and defined as $\Delta_{\text{IF-TEM}} = \frac{\kappa\delta}{(\alpha+1)(\alpha-1)} \frac{2}{cK}$. Incorporating this step size and the bound from Lemma \ref{thm0}, and considering the relationship between $E$, $\Omega$, and $c$ described earlier \eqref{eq:c_omega_connection}, we deduce the following upper bound
\begin{equation}
\EX[\varepsilon^2] < \frac{R}{(1-R)^2} \left(\frac{\alpha+1}{\alpha-1}\right)\left(\frac{E\Omega}{3\pi}\right) \frac{1}{2^{2N}},
\label{eq:31}
\end{equation}
where $R = \left(\frac{\kappa\delta}{\alpha-1}\right) \sqrt{\frac{\Omega}{E\pi}}<1$. Here, $R < 1$ since the IF-TEM parameters $\{\kappa, b,\delta\}$ are chosen to satisfy the Nyquist criterion \eqref{eq:density}, and achieve perfect recovery of the BL signal $x(t)$ from the IF-TEM time instances. This completes our proof.


\end{proof}



\begin{figure}[t!]
	\centering
	\hspace{-1.3cm}
	\includegraphics[width=0.45\textwidth]{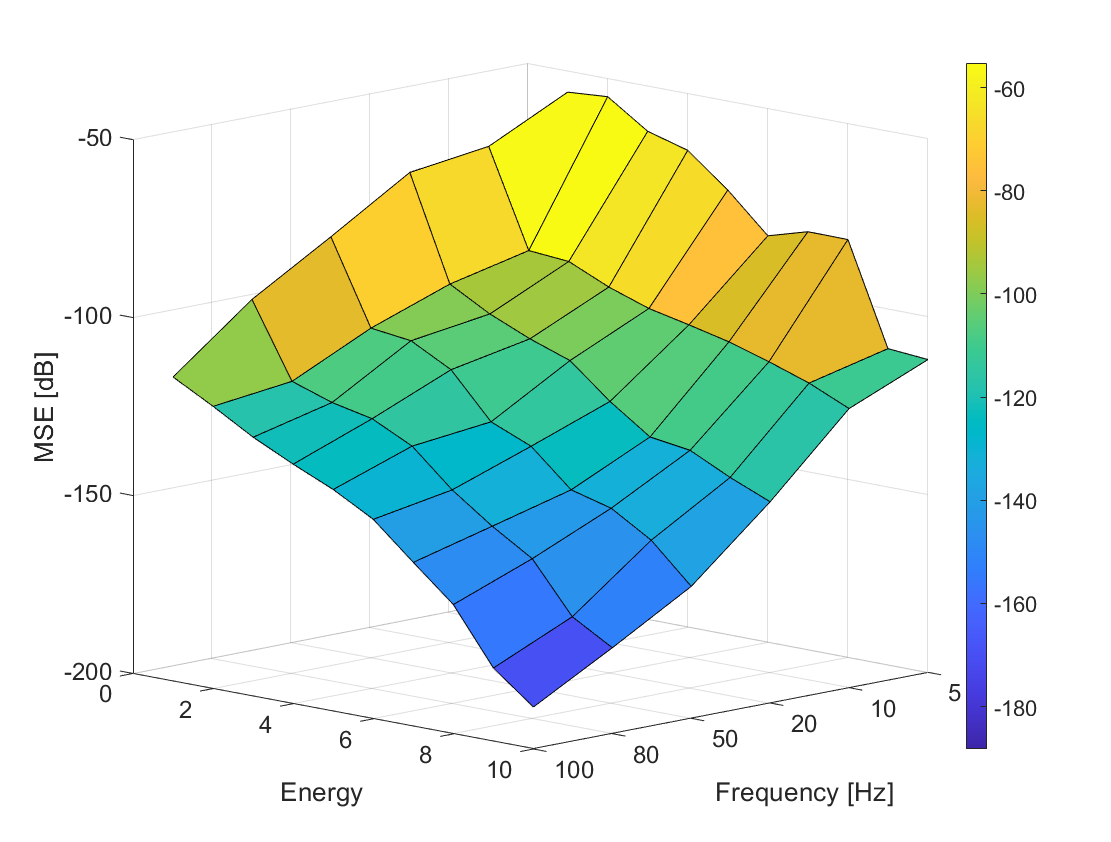}
	\caption{MSE using IF-TEM sampler as a function of the frequency and energy of the signal using 12 bits. Increasing frequency or energy decreases the MSE.}
	\label{fig:thm2}
\end{figure}

We present the results of our analysis in Fig. \ref{fig:thm2}, where we plot the MSE bound in decibels as a function of the number of bits and the signal frequency. Lemma \ref{thm0} is used to obtain these results, which do not take into account the interplay between the signal frequency and amplitude. As expected, the resolution and recovery of the signal improve with an increase in the total number of bits and bits per sample, respectively. However, an increase in the signal's energy or frequency leads to more time instances requiring quantization, resulting in a higher MSE.


To demonstrate the applicability of Theorem \ref{thm:Ebound}, we present in Fig. \ref{fig:thm2} the relationship between the MSE and the energy and frequency of a signal. As the energy and frequency of the signal increase, so does the total number of time instances $t_n$, which in turn results in a larger oversampling factor. Therefore, we select the IF-TEM sampler parameters such that a low oversampling factor is achieved while maintaining a good reconstruction in terms of MSE. The oversampling factor is defined as the ratio between the actual IF-TEM sampling rate and the Nyquist rate for a given bandwidth. 

As demonstrated in \cite{naaman2022fri}, the firing rate of an IF-TEM, $F_R$, with parameters $b, \kappa$, and $\delta$ is lower and upper bounded by
\begin{align}
   \frac{1}{\Delta t_{\max}} = \frac{b-c}{\kappa \delta} \leq F_R \leq  \frac{b+c}{\kappa \delta} = \frac{1}{\Delta t_{\min}}.
    \label{eq:firing_bounds}
\end{align}
Note that for a BL signal $x(t)$, $t\in\mathbb{R}$ defined over $[-\Omega,\Omega]$, the firing rate can be expressed as
\begin{equation}
    F_R = OS\cdot {f_{nyq}},
\end{equation}
where $OS$ is the oversampling factor and $f_{nyq} = \frac{\Omega}{\pi}$ is the Nyquist rate. Consequently, we have the relationship
\begin{equation}
   f_{nyq}\Delta t_{\min} \leq \frac{1}{OS}\leq f_{nyq}\Delta t_{\max}.
   \label{eq:OS_bounds}
\end{equation}

\begin{figure}[t!]
	\centering
	\hspace{-1.3cm}
	\includegraphics[width=0.45\textwidth]{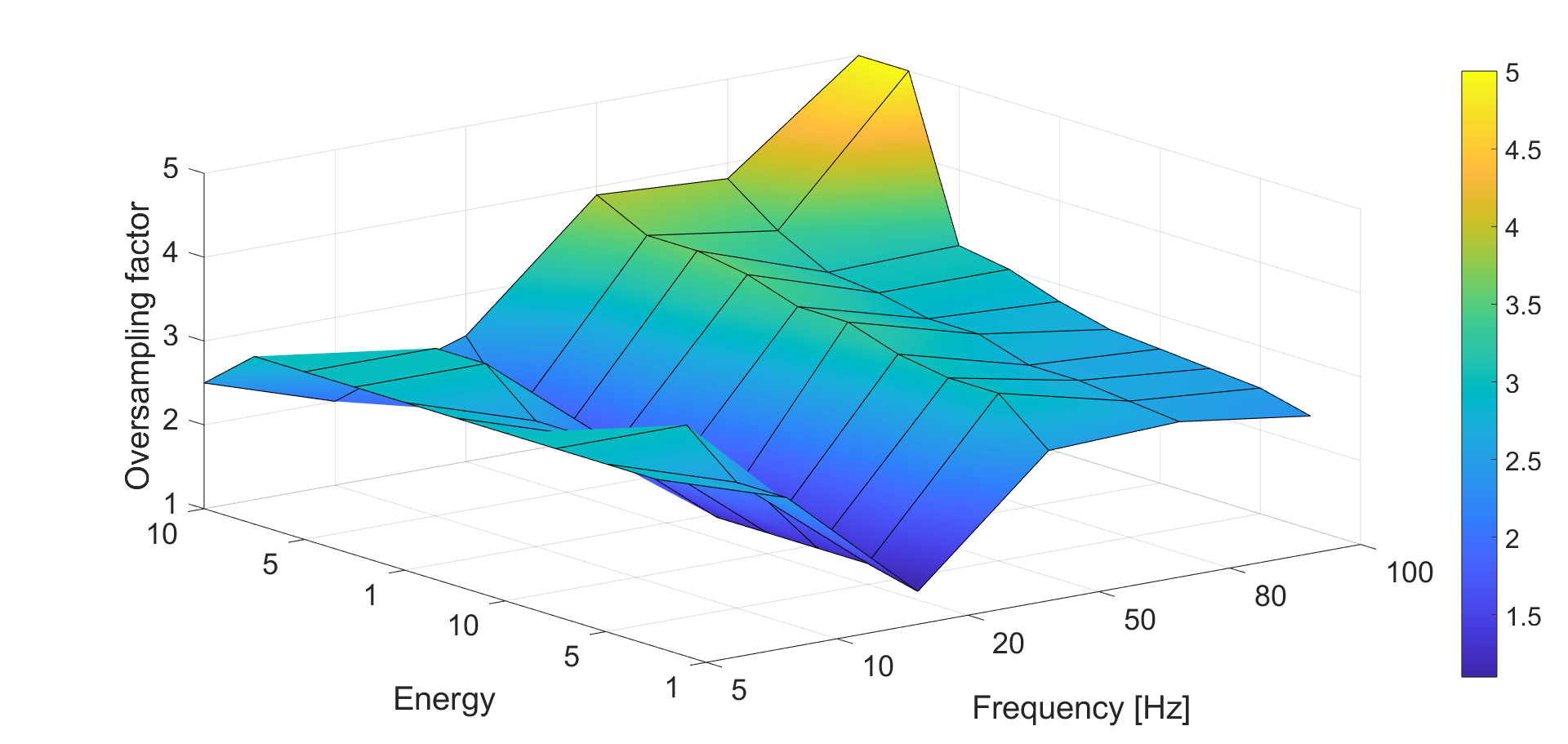}
	\caption{Oversampling factor as a function of energy and frequency. To fairly compare the MSE of the approaches in our simulations, we use the same amount of bits and a moderate oversampling factor.}
	\label{fig:thm22}
\end{figure}

The impact of signal frequency and energy on the quantization step size and oversampling ratio is demonstrated in Figs. \ref{fig:thm2} and \ref{fig:thm22}. Fig. \ref{fig:thm2} depicts the relationship between the MSE in decibels and the signal's energy and frequency. The plot shows that increasing the frequency and energy of the signal results in a reduction of the quantization step size and a decrease in the time instance discrepancies, leading to better signal reconstruction accuracy. Fig. \ref{fig:thm22} displays the frequency and energy dependence of the oversampling factor. As the frequency and energy of the signal increase, its maximal amplitude $c$ also increases, resulting in an increase in the oversampling ratio. These findings suggest that careful selection of the IF-TEM sampler parameters can achieve a low oversampling ratio while maintaining good signal reconstruction accuracy in terms of MSE.


\noindent
\subsection{Comparison with Classical Method}
We next compare the performance of conventional and IF-TEM reconstruction techniques in terms of their MSE. Our goal is to evaluate the advantages of using IF-TEM sampler over conventional sampling in the presence of quantization. 

Consider a uniform quantizer with $K=2^N$ levels and $N$ quantization bits. The quantization step size is determined using \eqref{eq:classic_stepsize}, which yields
\begin{equation*}
\Delta_{\text{classic}} = \frac{2c}{K}.
\end{equation*}
Here, $c$ denotes the maximal amplitude of the signal, and $K$ represents the total number of quantization levels. Assuming that the quantization error is a sequence of uniform i.i.d random variables on $[-\Delta_{\text{classic}}/2, \Delta_{\text{classic}}/2]$, the quantization error for the classical uniform sampler for the BL signal becomes \cite{tan2018digital}
\begin{equation}
\mathrm{MSE}_{\mathrm{classic}} = \frac{\Delta_{\text{classic}} ^2}{12} = \frac{(2c/K)^2}{12} = \frac{c^2}{3K^2},
\label{eq:NyqEq}
\end{equation}
where $\mathrm{MSE}_{\mathrm{classic}}$ denotes the quantization error. When oversampling the signal, the quantization error becomes 
\cite{tan2018digital}
\begin{equation}
\mathrm{MSE}_{\mathrm{classic}} = \frac{\Delta_{\text{classic}} ^2}{12}\cdot\frac{1}{OS} =\frac{c^2}{3K^2}\cdot\frac{1}{OS}.
\label{eq:NyqEq_oversampling}
\end{equation}
We use \eqref{eq:NyqEq_oversampling} for the quantization error of the classical signal when oversampling the signal.

In practical scenarios, both the classical sampler and the IF-TEM sampler operate over a finite time interval, resulting in a finite number of samples. To evaluate their performance, we calculate the MSE of the IF-TEM sampler over a sufficiently large time window, which closely approximates the expression in \eqref{eq:epsilon_square}.
The calculated MSE of the IF-TEM sampler over this large time window is bounded by the upper bounds derived in Theorem \ref{thm:Ebound}. The approximation using a sufficiently large time window enables a direct comparison of the quantization error matrics between classical ADC and IF-TEM sampler. Our aim is to compare the Nyquist ADC quantization error given by \eqref{eq:NyqEq_oversampling} to the IF-TEM MSE upper bound as given in \eqref{eq:UB_bits}. We summarize our findings in the following theorem.
\begin{theorem}
Let $x(t)$, $t\in\mathbb{R}$ be a signal BL to $[-\Omega,\Omega]$.
The signal $x(t)$ is sampled by an IF-TEM sampler and a classical uniform sampler with a fixed oversampling $OS$. Both the samplers are followed by a $K$-level uniform quantizer with $N = \log_2 K$ bits. Let $E$ be the maximal energy of $x(t)$, and let the relationship between $E$, $\Omega
$ and $c$ be given by $c = \sqrt{{E\Omega}/{\pi}}$. 
Let $b = \alpha c$ for any fixed  $\alpha>1$, where $b$ represents the IF-TEM bias.
A sufficient condition for IF-TEM to exhibit lower quantization noise than Nyquist ADC for a fixed number of bits $N$ is given by
\begin{equation}
    \left(\frac{1}{2(1-R)^2}\right)\left(\frac{\alpha+1}{\alpha-1}\right)^2\leq 1,
    \label{eq:cond_ob}
\end{equation}
where $\alpha=\frac{b}{c}>1$, and $R = \left(\dfrac{\kappa\delta}{\alpha-1}\right) \sqrt{\frac{\Omega}{E\pi}}<1$.
\label{thm:3}
\end{theorem}
\begin{proof}
Using \eqref{eq:UB_bits}, \eqref{eq:OS_bounds}, \eqref{eq:NyqEq_oversampling}, \eqref{eq:c_omega_connection} and $b = \alpha c$, where $\alpha>1$, it follows that 
\begin{align}
    \EX[\varepsilon^2] &< \dfrac{R}{(1-R)^2} \left(\dfrac{\alpha+1}{\alpha-1}\right)\left(\dfrac{E\Omega}{3\pi}\right) \dfrac{1}{2^{2N}}\nonumber\\
    &= \left(\dfrac{E\Omega}{3\pi}\right) \dfrac{1}{2^{2N}}\dfrac{1}{(1-R)^2} \left(\dfrac{\alpha+1}{\alpha-1}\right)\left(\dfrac{\kappa\delta}{\alpha-1}\right) \sqrt{\frac{\Omega}{E\pi}} \nonumber\\
    &= \left(\dfrac{E\Omega}{3\pi2^{2N}}\right)\dfrac{1}{(1-R)^2} \left(\dfrac{\alpha+1}{\alpha-1}\right)^2\left(\dfrac{\kappa\delta\Omega}{\pi c ({\alpha+1})}\right) \nonumber\\
    &= \left(\dfrac{E\Omega}{3\pi2^{2N}}\right)\dfrac{1}{(1-R)^2} \left(\dfrac{\alpha+1}{\alpha-1}\right)^2\left(\dfrac{F_{Nyq}}{2}\Delta t_{\min}\right) \nonumber\\
    &\leq \frac{1}{2}\cdot\underbrace{\frac{c^2}{3K^2}\frac{1}{OS}}_{\mathrm{MSE}_{\mathrm{classic}}}\cdot \dfrac{1}{(1-R)^2} \left(\dfrac{\alpha+1}{\alpha-1}\right)^2
    \label{eq:UBapprox}
\end{align}

To ensure that the IF-TEM has lower quantization noise than the Nyquist ADC, \eqref{eq:cond_ob} serves as a sufficient condition.

\end{proof}

\begin{figure}[!t]
\begin{center}
\begin{tabular}{cc}
\subfigure[Classical sampler]{\includegraphics[width = 1.7in]{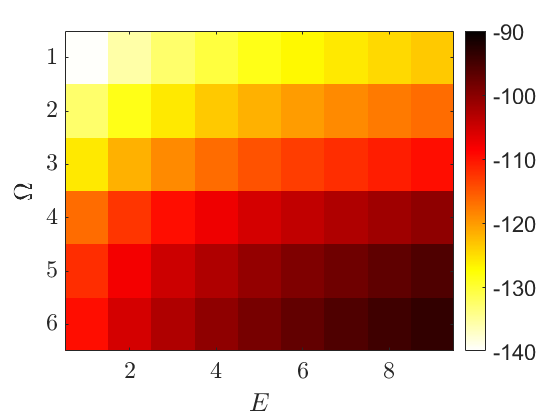}}&
\hspace{-.1in}\subfigure[IF-TEM sampler]{\includegraphics[width = 1.7in]{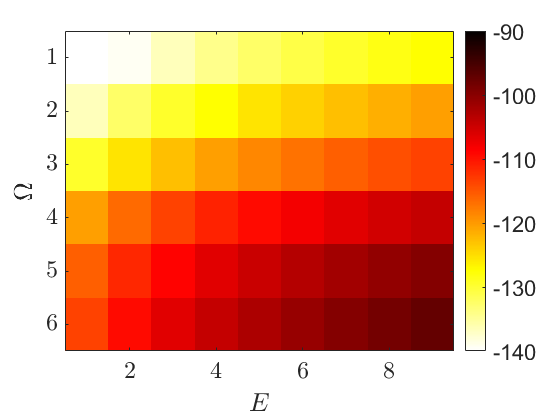}}\vspace{-.1in}
\end{tabular} 
\end{center}
\caption{A comparison between simulated MSE of the classical and IF-TEM samplers with constant oversampling.The MSE of IF-TEM is 8dB lower compared with that of the classical sampler.}
\label{fig:TE_compare}
\end{figure}

An illustration of the effectiveness of the chosen IF-TEM parameters in reducing the MSE compared to the classical sampler while maintaining constant oversampling is shown in Fig. \ref{fig:TE_compare}. Specifically, we consider a $2\Omega$ bandlimited signal $x(t)$ that is time-bounded, i.e., $|x(t)|\leq c$, where $c=\sqrt{(E\Omega)/\pi}$ with $E\in[2,10]$ and $\Omega$ fluctuating between $5-100$ Hz. 
The number of bits $N=8$.
IF-TEM method employs fixed values of $\kappa=0.4$ and $\alpha=b/c=10$. The threshold $\delta$ is selected to satisfy the requirement stated in equation \eqref{eq:cond_ob}, which yields an oversampling factor of 4.4. In this scenario, IF-TEM method yields an MSE that is 5 dB lower than that of the conventional method.

\begin{figure}[!t]
\begin{center}
\begin{tabular}{cc}
\subfigure[Classical sampler]{\includegraphics[width = 1.7in]{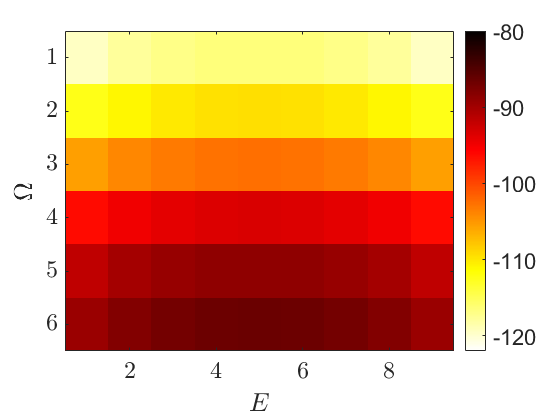}}&
\hspace{-.1in}\subfigure[IF-TEM sampler]{\includegraphics[width = 1.7in]{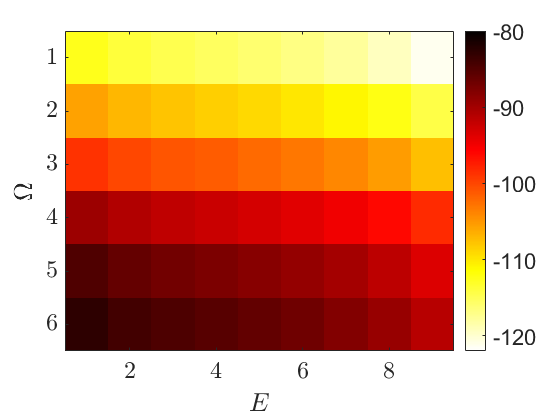}}\vspace{-.1in}
\end{tabular} 
\end{center}
\caption{A comparison of simulated MSE between the classical and IF-TEM samplers with a varying oversampling. With the increase of energy, the IF-TEM sampler has lower error compared to with the classical sampler.}
\label{fig:TE_compare2}
\end{figure}

Next, by permitting variable oversampling and evaluating the relationship in \eqref{eq:c_omega_connection}, as represented in Fig. \ref{fig:TE_compare2}, we demonstrate that as the signal energy grows, using IF-TEM sampler, the signal MSE drops. It is noteworthy that this relationship holds for the IF-TEM sampler, but not for the classical sampler, as can be seen from the figure.
Based on Theorem \ref{theorem:IF_Quant_BL}, we see that the quantization step size decreases as the energy of the signal increases. This decrease in quantization step size can result in a decrease in the MSE of the reconstructed signal. Therefore, as demonstrated in the example given in  Fig. \ref{fig:TE_compare}, the MSE can decrease as well as the energy of the signal increases.
The IF-TEM parameters are selected as follows; we use fixed values of $\kappa=2$, and $\alpha = \frac{b}{c}=5$. In order to have a sufficient number of samples required for recovery, the $\delta$ threshold is chosen to satisfy the requirement in \eqref{eq:cond_ob}. Here, $E\in[2,10]$, $\Omega$ vary from $5-100$ Hz, and the number of bits $N=8$.
As demonstrated, up to $E=4$, the classical sampler yields a lower MSE than the IF-TEM sampler. When $E>4$ and as the energy increases, the IF-TEM MSE becomes lower than the MSE of the conventional sampler.


Subsequently, we will prove a sufficient condition for the IF-TEM method to achieve a lower MSE than the classical ADC for the ratio of the IF-TEM bias $b$ to the amplitude bound $c$. The following theorem summarizes this result.

\begin{figure}[!t]
\begin{center}
\begin{tabular}{cc}
\hspace{-.2in}
\subfigure[Classical sampler]{\includegraphics[width = 3in]{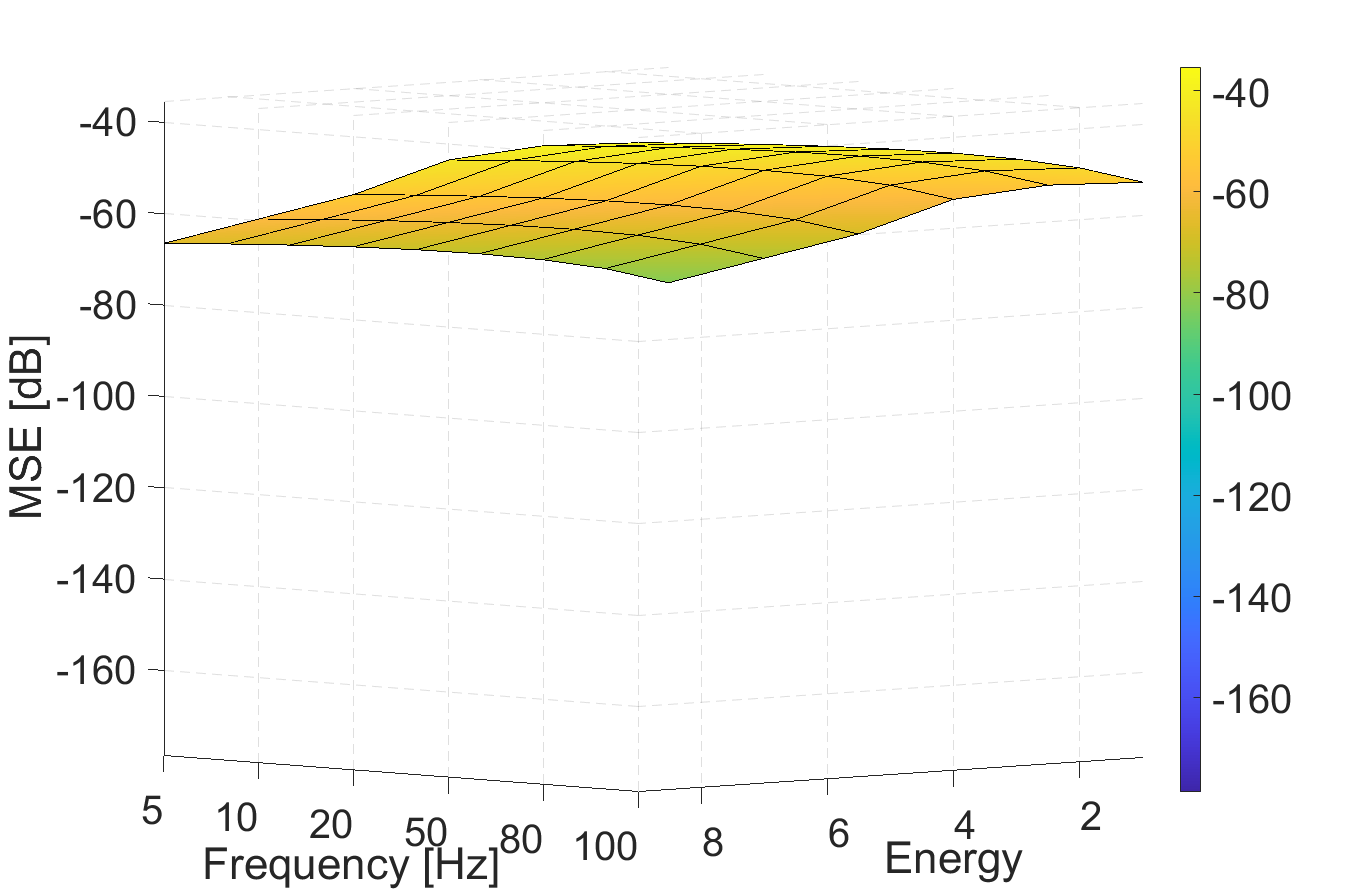}}\\\subfigure[IF-TEM sampler]{\includegraphics[width = 3in]{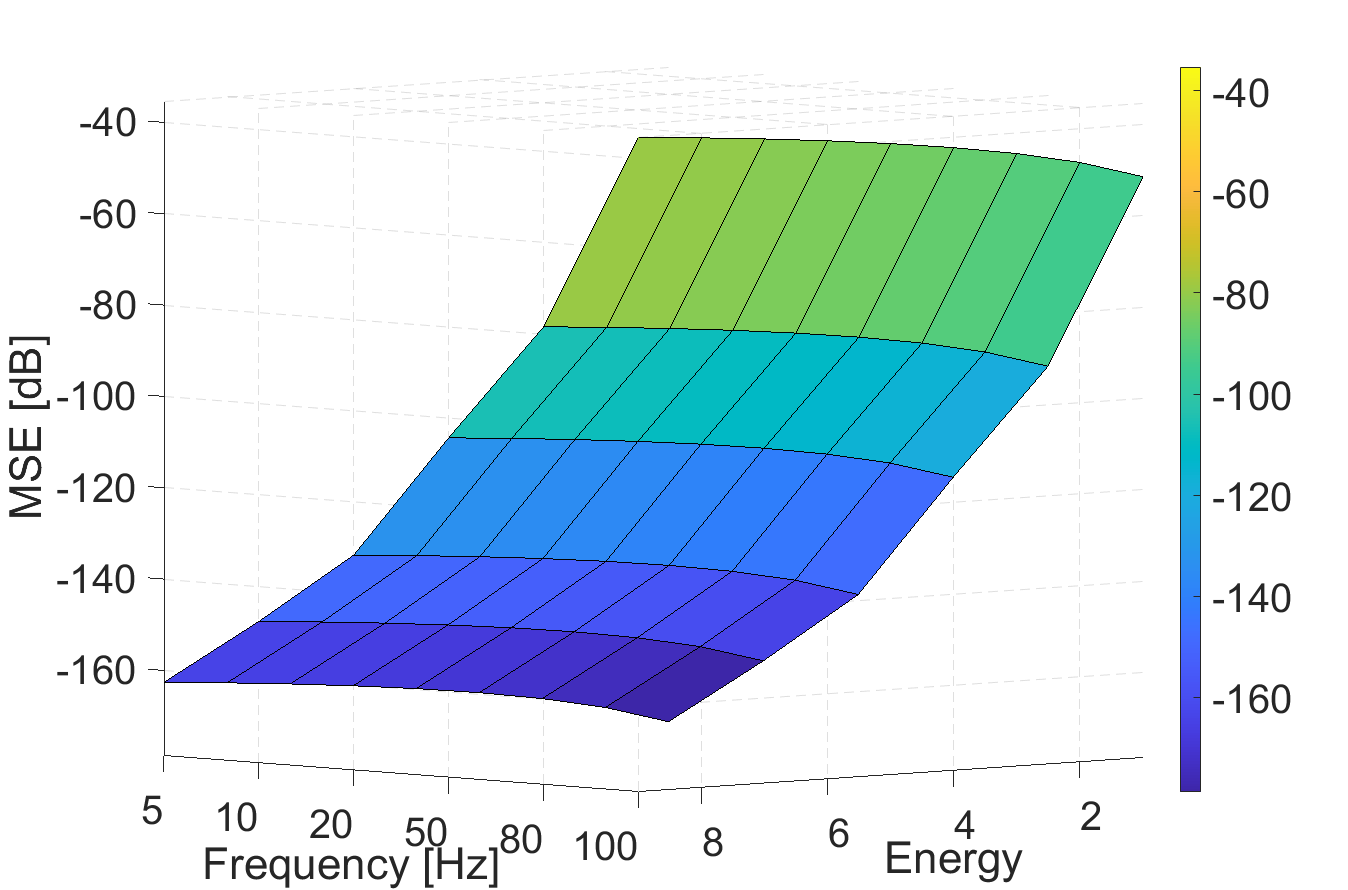}}\vspace{-.1in}
\end{tabular} 
\end{center}
\caption{Comparison of simulated MSE between (a) classical sampler and (b) IF-TEM ADC using 12 bits.}
	\label{fig:thm2_compare}
\end{figure}

\begin{theorem}
Let $x(t)$, $t\in\mathbb{R}$ be a signal BL to $[-\Omega,\Omega]$.
Let $E$ be the maximal energy of $x(t)$, and let the relationship between $E$, $\Omega
$ and $c$ be given by $c = \sqrt{{E\Omega}/{\pi}}$. 
Let $\alpha=\frac{b}{c}>1$, where $b$ is the IF-TEM bias. The IF-TEM achieves a lower MSE than the classical ADC if
\begin{equation}
\begin{split}
&\left(\alpha>1\right)  \quad\text{and} \quad  
    \left(\alpha \leq \left(3+\beta - \sqrt{\beta^2+6\beta+6}\right)\right)  \quad\text{or} \\&\quad \left(\alpha \geq\left(3+\beta + \sqrt{\beta^2+6\beta+6}\right)\right),
\end{split}
\end{equation}
where $\beta = \kappa\delta\sqrt{\frac{\Omega}{E\pi}}$.
\label{thm:4}
\end{theorem}
\begin{proof}
See Appendix \ref{app:thm4}.
\end{proof}
Moreover, that the condition in Theorem \ref{thm:4} is sufficient but not a necessary condition. 
Note the conditions $\alpha>1$ with $\left(\alpha \geq\left(3+\beta + \sqrt{\beta^2+6\beta+6}\right)\right)$ always yields a possible selection of $\alpha$ resulting in the IF-TEM achieving a lower MSE than the classical ADC.
 A significant observation arises from the fact that while this sufficient condition impacts the bias selection, it also plays a role in determining the sampling rate of the IF-TEM. This is particularly significant due to the Nyquist criterion $R = \left(\frac{\kappa\delta}{\alpha-1}\right) \sqrt{\frac{\Omega}{E\pi}}<1$. To closely approach the Nyquist rate, it necessitates $R\approx1$, which means that the IF-TEM parameters ${\kappa,\delta}$ are selected in such a way that ensures $\kappa\delta \approx \sqrt{\frac{E\pi}{\Omega}}(\alpha-1)$.

Fig. \ref{fig:thm2_compare} presents a comparison between IF-TEM sampler and the classical sampler satisfying the conditions outlined in Theorem \ref{thm:4}.
The maximal signals amplitude $c$ is defined in \eqref{eq:cenergy}, $\Omega\in\{5,10,20,50,80,100\}$, and $E\in[1,10]$ satisfy \eqref{eq:c_omega_connection}. The parameters $\{\kappa,\delta\}$ are chosen as $\kappa=0.6$ and $\delta$ varying between $(0.6,6)$. The bias $b$ is selected such that $b=\alpha c$ with $\alpha=8$. Both the IF-TEM and classical sampler employ an equal number of $N=12$ bits and an identical total number of samples with an OS of up to $6$. It is observed that the MSE decreases as the frequency and energy of the signal increase for the IF-TEM sampler, while there is no such trend for the classical sampler.

In the next section, we show that similar observations are true for the class of FRI signals.

\section{IF-TEM for FRI signals}
\label{sec:FRI}
In this section, we analyze quantization strategies for classical and IF-TEM sampling schemes with FRI signals. 
We show that as the number of pulses $L$ increases for FRI signals, the dynamic range of each sample decreases. We, therefore, suggest increasing the resolution of the quantizer as a function of $L$.

Similar to the case of BL signals, the sampled signal is quantized by uniform scalar quantizer with a resolution of $\log_2 K$ bits, i.e., the quantizer produce $K$ distinct output values.  
For FRI signals, given that the SoS filter is bounded, the sampler input $y(t)$ is also bounded
\cite{naaman2022fri,papoulis1967limits} 
\begin{align}
    |y(t)| \leq c = 
{L\,\,a_{\max} \,\,
    \|g\|_{\infty} \|h\|_{1}}.
    \label{eq:c2}
\end{align}
For the IF-TEM sampler, we quantize the time-differences $T_n$. The IF-TEM step-size is given by \eqref{eq:iftem_stepsize}.
For FRI signals recovery from IF-TEM sampler, using \eqref{eq:FRI_rec} requires a number of samples $N\geq 2L+2$.
When increasing $L$, we can increase the bias $b$ or decrease the threshold $\delta$ to have a sufficient number of samples for recovery. 
To analyze the relation between $L$ and $\Delta_{\text{IF-TEM}}$, fixed values of $\delta$ and $\kappa$ are assumed, while $b$ changes.
We show that by increasing $L$, the quantization step size $\Delta_{\text{IF-TEM}}$ decreases.
We summarize this result in the following theorem.  

\begin{theorem}
\label{theorem:IF_Quant}
{Consider an IF-TEM sampler followed by a K-level uniform quantizer with $N$ bits, i.e., $K=2^N$. Given a fixed $\alpha > 1$, let the IF-TEM bias be represented by $b$, such that $b = \alpha c$, where $c$ denotes the maximal signal amplitude.}
For FRI signals, the quantization step $\Delta_{\text{IF-TEM}}$ decreases as the number of input pulses $L$ increases.
\end{theorem}

\begin{proof}
Fix $\kappa$ and $\delta$. The bias is chosen such that $b = \alpha c$ with fixed $\alpha >1$. Substituting $b$ into \eqref{eq:iftem_stepsize}, we have
\begin{equation}
    \Delta_{\text{IF-TEM}} = \frac{\kappa\delta}{(\alpha+1)(\alpha-1)} \frac{2}{cK}.
    \label{eq:iftem_stepsize2}
\end{equation}
Using \eqref{eq:FRI_rec} and \eqref{eq:c2} conditions, with an increasing number of pulses $L$, the IF-TEM quantization step size will decrease.
\end{proof}

For a fixed $K$, increasing $L$ for an FRI signal increases the number of samples $N$ in both IF-TEM and classical methods.  Thus, the total number of bits will be increased in both cases. 

Note that FRI signals are determined by a finite number $L$ of unknowns, referred to as innovations, per time interval $T$. BL signals, have $L=1$ innovations per Nyquist interval $T=\frac{1}{f_{nyq}}=\frac{1}{2\Omega}$. Thus, increasing $\Omega>0$ means decreasing $T$, which causes a similar effect of reducing the quantization step size to increase $L$.  The time instances become closer, which causes smaller values of $T_n$s.
Thus, the quantization error can be reduced based on dense quantization, and the IF-TEM framework results in lower quantization error than the classical scheme.

\begin{figure}[t!]
\centering
\includegraphics[width=0.5\textwidth]{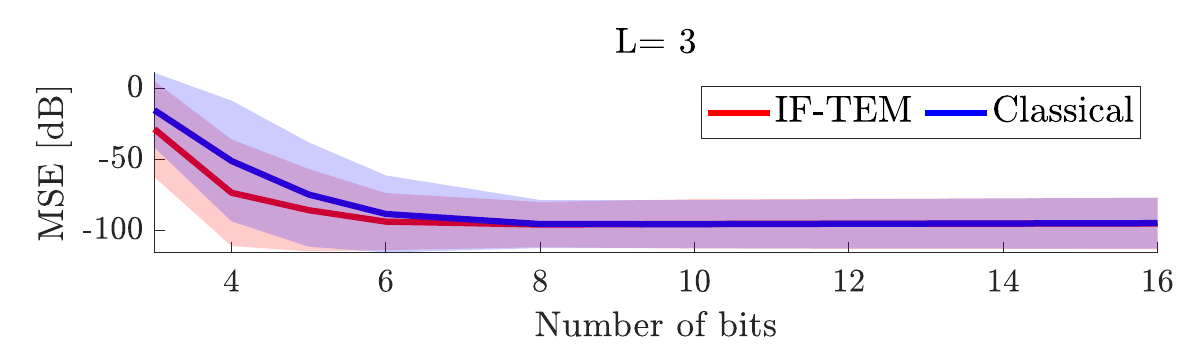} \\
\includegraphics[width=0.5\textwidth]{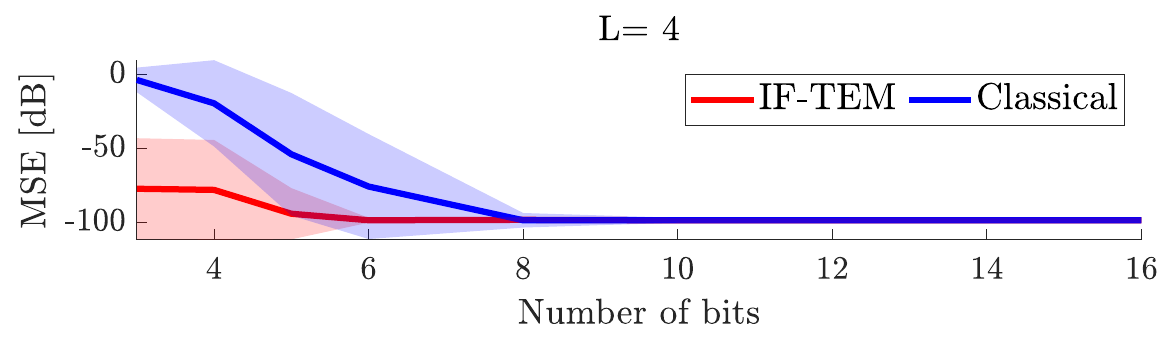} \\
\label{fig:1}
\includegraphics[width=0.5\textwidth]{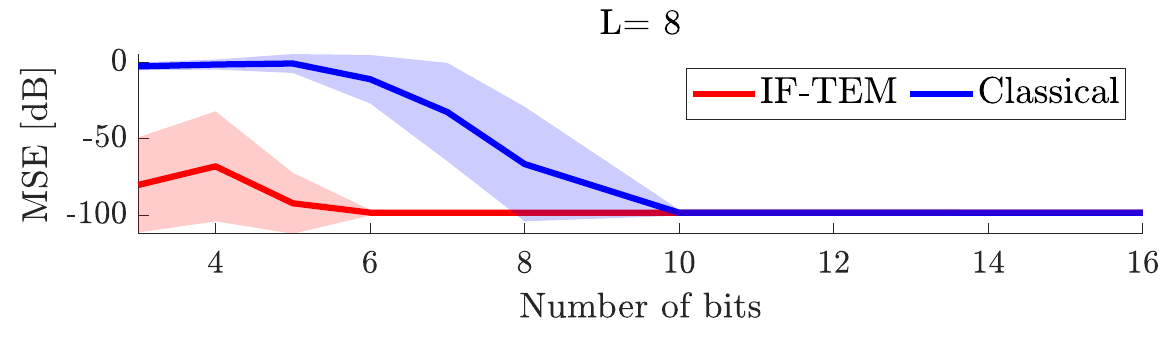}  
\caption{Mean-squared error in estimating an FRI signal as a function of the number of bits. The solid line shows the average error and the shaded region captures the standard deviation in the estimate.}
        \vspace{-0.5cm}
        \label{fig:yyrec}
\end{figure}

In Fig.~\ref{fig:yyrec}, we numerically demonstrate Theorem \ref{theorem:IF_Quant}.
The suggested IF-TEM sampling framework with quantization is then evaluated in terms of MSE and compared to the conventional approach using an FRI signal model.
In particular, we consider an FRI signal $x(t)$ as in \eqref{eq:fri}, with period $T = 1$ seconds which consists of $L=3$, $L=4$, and $L=8$ impulses, with 500 randomly selected amplitudes within the range  $[-1,1]$. The time-delays are selected randomly within the range $(0,1]$ with a resolution grid of $0.05$. For both the classical and IF-TEM FRI schemes, we consider an SoS sampling kernel that aids in selecting $2L$ FSCs \cite{naaman2022fri}.
For each signal $|x(t)|\leq c$, where $c$ is defined in \eqref{eq:c2}, the IF-TEM parameters are chosen as follows: $b = 10c$, $\delta = 30$, and $\kappa\in \{0.5,2\}$ for $L=3,4$ and $L=8$ respectively (without any quantization the error is -98.8 dB). The number of samples is the same for each data point in the classical and IF-TEM schemes and is approximately $8L$. After computing the FSCs of $x(t)$, the FRI parameters are computed by applying orthogonal matching pursuit to both classical and IF-TEM methods \cite{eldar2015sampling}. 
Reconstruction accuracy of the two methods is compared in terms of MSE, given by 
\begin{equation}
    \text{MSE} = \frac{||{x(t)-\Bar{x}(t)}||_{L_2[0,T]}}{||{x(t)}||_{L_2[0,T]}},
\end{equation}
where $\Bar{x}(t)$ is the reconstructed signal.

In Fig. \ref{fig:yyrec}, a comparison between the MSE of the recovered signals from the IF-TEM sampler (in red) and the classical sampler (in blue) is shown. In the IF-TEM, the difference between the time instances is quantized, whereas, in the conventional method, the amplitudes are quantized.
For each data point, the same number of samples and bits are used.
First, as shown in Fig. \ref{fig:yyrec}, using the IF-TEM sampler results in MSE reduction of at least 5dB less using up to 8 bits, compared to the classical sampler.
When the number of bits is greater than 8, almost perfect recovery is achieved in both methods.
Second, when increasing the number of pulses $L$, or raising the rate of innovation, the MSE is further decreased. 
As increasing the number of pulses for FRI signals is similar to increasing the input signal frequency for BL signals, the same behaviour holds for BL signals. Deriving error bounds for signal recovery by quantizing IF-TEM samples for FRI signals, as compared to classical uniform samplers in terms of MSE, is beyond the scope of this paper.

\section{Conclusion}
\label{sec:conc}
In this work, we analyzed the effects of quantization on IF-TEM sampler and demonstrated its advantages over classical ADC. Specifically, we show that increasing the bandwidth of a BL signal or the number of pulses of an FRI signal allows us to reduce the quantization step size when the number of quantization bits is fixed. An upper bound on the signal recovery error is derived for BL signals. Our theoretical and experimental results indicate that, with the same number of quantization bits, the IF-TEM sampler with quantized time difference measurements can achieve lower MSE than uniform samplers with uniform amplitude quantization for both BL and FRI signal models.



\appendix
\subsection{Proof of Lemma 1}
\label{app:prop1}
Let $x(t)$ be a BL signal with bandwidth $[-\Omega,\Omega]$ and maximum amplitude $c$. Consider an IF-TEM sampler followed by a $K$-level uniform quantizer with $N = \log_2 K$ bits.
The quantization error $d_k = \hat{T}_k -T_k$ is a sequence of uniform i.i.d random variables on $[-\Delta_{\text{IF-TEM}}/2, \Delta_{\text{IF-TEM}}/2]$
The IF-TEM parameters $\{\kappa, b>c,\delta\}$ are chosen to satisfy the Nyquist criterion \eqref{eq:density}, i.e., $r \triangleq \dfrac{\kappa\delta\Omega}{\pi (b-c)}<1$.
As a result, the difference between any two consecutive values of the measured time sequence $\{\hat{t}_k\}$ is bounded by the inverse of the Nyquist rate, i.e., $\sup_k{\hat{T}_k} < T$.
Let $w_n(t)$ be defined as
\begin{equation}
    w_n(t) = \mathds{1}_{[-n\Delta t_{\min},n\Delta t_{\min}]}.
\end{equation}
In this case, the MSE is calculated using \eqref{eq:e} and \eqref{eq:epsilon_square}, which yields
    \begingroup
    \allowdisplaybreaks
\small{
\begin{align}
    \varepsilon^2  & = \lim_{n\rightarrow\infty} \frac{1}{2n\Delta t_{\min}} {||e \mathds{1}_{[-n\Delta t_{\min},n\Delta t_{\min}]}||}^2 \nonumber\\
     & =
    \lim_{n\rightarrow\infty}\frac{1}{2n\Delta t_{\min}}{||\sum_{k\in Z}(I-\hat{\mathcal{R}})^k \sum_{l\in Z}\epsilon_l \sinc_\Omega(t-\hat{s}_l) w_n(t)||}^2\nonumber \\
    &\leq
    {||\sum_{k\in Z}(I-\hat{\mathcal{R}})^k||}^2 \lim_{n\rightarrow\infty} \frac{1}{2n\Delta t_{\min}}{||\sum_{l\in Z}\epsilon_l \sinc_\Omega(t-\hat{s}_l) w_n(t)||}^2 \nonumber\\
    &\leq
    \frac{1}{(1-r)^2} \lim_{n\rightarrow\infty} \frac{1}{2n\Delta t_{\min}}{||\sum_{l\in Z}\epsilon_l \sinc_\Omega(t-\hat{s}_l) w_n(t)||}^2.
    \label{epsilonLEQ}
\end{align}
}
\endgroup
This inequality serves as an upper bound for $\varepsilon$ the error signal. Thus, the MSE is upper bounded by
\begin{equation}
\begin{split}
    \EX[\varepsilon^2] &\leq \EX \left[ \frac{1}{(1-r)^2} \lim_{n\rightarrow\infty} \frac{1}{2n\Delta t_{\min}}{||\sum_{l\in Z}\epsilon_l \sinc_\Omega(t-\hat{s}_l) w_n(t)||}^2 \right]\\
    &= \frac{1}{(1-r)^2} \lim_{n\rightarrow\infty}  \frac{1}{2n\Delta t_{\min}}\EX\left[{||\sum_{l\in Z}\epsilon_l \sinc_\Omega(t-\hat{s}_l) w_n(t)||}^2 \right].
    \label{eq:mseleq}
  \end{split}
\end{equation}
Next, we bound the expectation on the right-hand side as follows
\begingroup
\allowdisplaybreaks
\begin{align}
    &\EX\left[{||\sum_{l\in Z}\epsilon_l \sinc_\Omega(t-\hat{s}_l) w_n(t)||}^2 \right]  \nonumber\\
 &=\int_{-n \Delta t_{\min}}^{n \Delta t_{\min}}{\sum_{m\in Z}\sum_{k\in Z} \sinc_\Omega(t-\hat{s}_m) \sinc_\Omega(t-\hat{s}_k)}dt \EX[\epsilon_k \epsilon_m]\nonumber\\
     &=\int_{-n \Delta t_{\min}}^{n \Delta t_{\min}}{\sum_{k\in Z} {\sinc_\Omega(t-\hat{s}_k)}^2 dt} \left(\frac{\kappa\delta}{T_k}\right)^2\frac{\Delta_{\text{IF-TEM}}^2}{12},
\end{align}
\endgroup
where it can be inferred from Appendix \ref{res:3} that
\begin{equation}
     \EX[\epsilon_k \epsilon_m] = \left(\frac{\kappa\delta}{T_k}\right)^2 \frac{\Delta_{\text{IF-TEM}} ^2}{12} \delta_{k,m}.
     \label{eq:epsiloneq}
\end{equation}
Given that 
\begin{equation}
    \Delta t_{\min} = \frac{\kappa\delta}{b+c},
    \label{eq:Tmin}
\end{equation}
we can deduce that
\begin{align}
    &\EX\left[{||\sum_{l\in Z}\epsilon_l \sinc_\Omega(t-\hat{s}_l) w_n(t)||}^2 \right]\nonumber   \\
    &\leq \int_{-n \Delta t_{\min}}^{n \Delta t_{\min}}{\sum_{k\in Z} {\sinc_\Omega(t-\hat{s}_k)}^2 dt} \left({\frac{\kappa\delta}{\Delta t_{\min}}}\right)^2\frac{\Delta^2}{12}\nonumber\\
    & =\int_{-n \Delta t_{\min}}^{n \Delta t_{\min}}{\sum_{k\in Z} {\sinc_\Omega(t-\hat{s}_k)}^2 dt} (b+c)^2\frac{\Delta_{\text{IF-TEM}}^2}{12}.
  \end{align}
Furthermore, as given in  \cite[Result 3]{lazar2004perfect}:
\begin{equation}
    \frac{1}{2n} \int_{-n \Delta t_{\min}}^{n \Delta t_{\min}}{\sum_{k\in Z} {\sinc_\Omega(t-\hat{s}_k)}^2 dt} \leq \frac{1}{T} 
\end{equation}
Therefore, 
\begin{equation}
\begin{split}
    \EX\left[{||\sum_{l\in Z}\epsilon_l \sinc_\Omega(t-\hat{s}_l) w_n(t)||}^2 \right] &\leq \frac{2n}{T} (b+c)^2 \frac{\Delta^2}{12} .
    \label{eq:partialresult}
  \end{split}
\end{equation}
Using \eqref{eq:mseleq}, \eqref{eq:Tmin} and \eqref{eq:partialresult} results
\begingroup
\allowdisplaybreaks
\begin{align}
    \allowdisplaybreaks
    \EX[\varepsilon^2] 
    &\leq \frac{1}{(1-r)^2} \lim_{n\rightarrow\infty}  \frac{\EX\left[{||\sum_{l\in Z}\epsilon_l \sinc_\Omega(t-\hat{s}_l) w_n(t)||}^2 \right]}{2n\Delta t_{\min}}\nonumber \\ &
    \leq \frac{1}{(1-r)^2} \lim_{n\rightarrow\infty}  \frac{1}{2n\Delta t_{\min}}\frac{2n}{T}(b+c)^2 \frac{\Delta_{\text{IF-TEM}}^2}{12} \nonumber\\ &\leq \left(\frac{b+c}{1-r}\right)^2 \left(\frac{b+c}{\kappa\delta T}\right) \frac{\Delta_{\text{IF-TEM}}^2}{12},
\end{align}
\endgroup
which completes the proof.
\subsection{Proof of Equation \eqref{eq:epsiloneq}}
\label{res:3}
We start by expressing $\epsilon_k$ as
\begingroup
\allowdisplaybreaks
\begin{align*}
\epsilon_k &= [\kappa\delta - b\hat{T}_k] - \int_{\hat{t}_k}^{\hat{t}_{k+1}} x(u)\mathrm{d}u\\
&= \int_{t_k}^{t_{k+1}}x(u)du - \int_{\hat{t}_k}^{\hat{t}_{k+1}} x(u)\mathrm{d}u-bd_k,
\end{align*}
\endgroup
where $d_k = \hat{T}_k-T_k$. By applying the mean-value theorem, we obtain
\begin{equation}
    \epsilon_k = x(\zeta_k)T_k-x(\hat{\zeta}_k)\hat{T}_k -bd_k,
\end{equation}
where $\zeta_k\in(t_k,t_{k+1})$ and $\hat{\zeta}_k\in(\hat{t}_k,\hat{t}_{k+1})$.
Note that the quantized IF-TEM values are denoted as $\hat{T}_k = T_k + d_k$, where $d_k$ is the quantization error, The recovered time encodings $\hat{t}_k$ can be expressed as $\hat{t}_k = \sum_{i=1}^k \hat{T}_i = \sum_{i=1}^k T_i+\sum_{i=1}^k d_i$. Assuming that $d_k $ uniform i.i.d random variables on $[-\Delta_{\text{IF-TEM}}/2, \Delta_{\text{IF-TEM}}/2]$, the variance of the error in 
$\hat{t}_k$ grows with $k$, given by, $\Var(t_k-\hat{t}_k) = \frac{k\Delta_{\text{IF-TEM}}^2}{12}$. This implies that the errors propagate to later measurements.
However, relying on Appendix \ref{res:2}, we can choose $\Delta_{\text{IF-TEM}}$ to be sufficiently small such that the following term is finite and bounded
\begin{equation}
    \max\limits |t_k-\hat{t}_k| \leq \frac{k\Delta_{\text{IF-TEM}}}{2}.
    \label{eq:small_delta}
\end{equation}
Hence, the variance of the difference between $t_k$ and $\hat{t}_k$ remains bounded, demonstrating its non-explosive nature. In particular, this property holds for any $\zeta_k\in(t_k,t_{k+1})$, as the signal $x(t)$ exhibits continuity. By leveraging the continuity property, we can approximate $\zeta_k \approx\hat{\zeta}_k$.


Let us denote $a_n$ to be
\begin{equation}
\begin{split}
     a_n =& \frac{1}{(1-r)^2} \lim_{n\rightarrow\infty}  \frac{1}{2n\Delta t_{\min}}\\ &\cdot\EX\left[{||\sum_{l\in Z}\epsilon_l \sinc_\Omega(t-\hat{s}_l) w_n(t)||}^2 \right].
\end{split}
\end{equation}
We have proven in Appendix \ref{res:2} that 
\begin{equation}
     \lim_{n\rightarrow\infty} \frac{1}{2n\Delta t_{\min}}{||\sum_{l\in Z}\epsilon_l \sinc_\Omega(t-\hat{s}_l) w_n(t)||}^2
     \label{eq:conv}
\end{equation}
converges. According to the definition of the limit, we have
\begin{equation}
    \forall \epsilon>0\exists n^{\prime}\in\mathbb{N}\quad \text{s.t.}\quad -\epsilon\leq a_{n^{\prime}} - \lim\limits_{n\rightarrow\infty} a_n \leq \epsilon
\end{equation}
since $\EX[\varepsilon^2]\leq \lim\limits_{n\rightarrow\infty} a_n\leq a_{n^{\prime}}+\epsilon$.
Using this ${n^{\prime}}$, we can choose $\Delta_{\text{IF-TEM}}$ to be sufficiently small such that
\begin{equation}
    \max_k \limits |t_k-\hat{t}_k| \leq \frac{k\Delta_{\text{IF-TEM}}}{2}\leq \frac{(2n^\prime+1)\Delta_{\text{IF-TEM}}}{2}.
\end{equation}
This implies that in this case, $\zeta_k\approx\hat{\zeta}_k$, and we can approximate $\epsilon_k$ as $\epsilon_k \approx \left(-x(\zeta_k)-b\right)d_k$.
Since 
\begin{equation}
    x(\zeta_k) = \frac{1}{T_k}\int_{t_k}^{t_{k+1}}x(u)du = -b+\frac{\kappa\delta}{T_k},
\end{equation}
we have $\epsilon_k \approx \frac{\kappa\delta}{T_k}d_k$.
Therefore,
\begin{equation}
     \EX[\epsilon_k \epsilon_m] = \frac{\kappa\delta}{T_k}\frac{\kappa\delta}{T_m}\EX[d_k d_m] =\left(\frac{\kappa\delta}{T_k}\right)^2 \frac{\Delta_{\text{IF-TEM}} ^2}{12} \delta_{k,m},
\end{equation}
which completes the proof.
\subsection{Proof of the convergence of \eqref{eq:conv}}
\label{res:2}
To prove convergence of \eqref{eq:conv}, we first consider the boundedness and increasing nature of the term
\begin{equation}
   \EX \left[ {||\sum_{l\in Z}\epsilon_l \sinc_\Omega(t-\hat{s}_l) w_n(t)||}^2 \right].
\end{equation}
We begin by demonstrating the boundedness of this term
\begingroup
\allowdisplaybreaks
\begin{align}
   &\EX \left[ {||\sum_{l\in Z}\epsilon_l \sinc_\Omega(t-\hat{s}_l) w_n(t)||}^2 \right]\nonumber\\
   & \leq \EX \left[ {\sum_{l\in Z}|\epsilon_l|^2 ||\sinc_\Omega(t-\hat{s}_l) w_n(t)||}^2 \right]\nonumber\\
   &\leq \max\limits_l |\epsilon_l|^2 \left[ {\sum_{l\in Z} \EX||\sinc_\Omega(t-\hat{s}_l) w_n(t)||}^2 \right]\nonumber\\
   &\leq \max\limits_l |\epsilon_l|^2 ||\sinc_\Omega(t)||_2^2 |{\#}l|,
\end{align}
\endgroup
where we note that as shown in Appendix \ref{res:1} $\epsilon_l$ is bounded, and $|\# l|$ is the number of spikes (time instances) in the window. Note that the number of spikes in a window in $[-n\Delta t_{\min},n\Delta t_{\min}]$, is upper bounded by $\frac{2n\Delta t_{\min}}{\Delta t_{\min}}+1 = 2n+1$. Thus, we have
\begin{equation}
    |{\#}l|\leq 2n+1.
\end{equation}
and consequently,
\begin{equation}
\begin{split}
       &\EX \left[ {||\sum_{l\in Z}\epsilon_l \sinc_\Omega(t-\hat{s}_l) w_n(t)||}^2 \right]\\&\leq \left(\kappa\delta + \left(b+c\right)\left( \Delta t_{\max} + \frac{\Delta_{\text{IF-TEM}}}{2}\right)\right)^2||\sinc_\Omega(t)||_2^2(2n+1).
\end{split}
\end{equation}
Moreover, the above norm is shown to be increasing in $n$ \cite{lazar2004perfect}. 
Dividing $\EX \left[ {||\sum_{l\in Z}\epsilon_l \sinc_\Omega(t-\hat{s}_l) w_n(t)||}^2 \right]$ by the window size $2n\Delta t_{\min}$, we find that its upper bound is a constant. Therefore, we have established that the following term converges:
\begin{equation*}
    \lim_{n\rightarrow\infty} \frac{1}{2n\Delta t_{\min}}{||\sum_{l\in Z}\epsilon_l \sinc_\Omega(t-\hat{s}_l) w_n(t)||}^2,\\
\end{equation*}
which completes the proof.
\subsection{Proof of boundedness of \eqref{eq:eps}}
\label{res:1}
Here we establish the boundedness of $\epsilon_l$ as defined in \eqref{eq:eps}.
Taking into account the definition provided in \eqref{eq:eps},
we prove that $\epsilon_l$ is bounded.
The bound on the absolute value of $\epsilon_l$ is evaluated as
\begin{equation}
    \begin{split}
        |\epsilon_l| &=\Bigg| [\kappa\delta - b\hat{T}_l] - \int_{\hat{t}_l}^{\hat{t}_{l+1}} x(u)\mathrm{d}u\Bigg|\\&\leq \kappa\delta + b|\hat{T}_l|+\Bigg | \int_{\hat{t}_l}^{\hat{t}_{l+1}} x(u)\mathrm{d}u\Bigg |.\\
    \end{split}
    \label{eq:69}
\end{equation}
Considering that $\hat{T}_l = T_l + d_l$, where $\Delta t_{\min}\leq T_l \leq \Delta t_{\max}$ and the quantization error $d_l \in [-\frac{\Delta_{\text{IF-TEM}}}{2},\frac{\Delta_{\text{IF-TEM}}}{2}]$, we can derive
\begin{equation}
    |\hat{T}_l|\leq | T_l| + | d_l|\leq \Delta t_{\max} + \frac{\Delta_{\text{IF-TEM}}}{2}.
\end{equation}
Furthermore, employing the mean-value theorem, the integral term can be expressed as
\begingroup
\allowdisplaybreaks
\begin{align}
     \Bigg | \int_{\hat{t}_l}^{\hat{t}_{l+1}} x(u)\mathrm{d}u\Bigg |&=|x(\zeta_l)||\hat{t}_{l+1}-\hat{t}_l|\nonumber\\
    &\leq c|\hat{T}_l|\leq c\left( \Delta t_{\max} + \frac{\Delta_{\text{IF-TEM}}}{2}\right),
    \label{eq:71}
\end{align}
\endgroup
where $\zeta_l\in (\hat{t}_l,\hat{t}_{l+1})$ and $x(t)\leq c$.
Hence, by applying \eqref{eq:71} in \eqref{eq:69}, we can conclude that
\begin{equation}
    |\epsilon_l|\leq \kappa\delta + \left(b+c\right)\left( \Delta t_{\max} + \frac{\Delta_{\text{IF-TEM}}}{2}\right),
\end{equation}
which completes the proof.

\subsection{Proof of Theorem 4}
\label{app:thm4}
Based on Theorem \ref{thm:3}, we can deduce that $0<R<1$ and that $\left(\frac{1}{2(1-R)^2}\right)\left(\frac{\alpha+1}{\alpha-1}\right)^2\leq1$, where $\alpha=\frac{b}{c}>1$, and $R = \left(\dfrac{\kappa\delta}{\alpha-1}\right) \sqrt{\frac{\Omega}{E\pi}}<1$. This inequality holds if and only if $0<R<1$ and $(1-R)^2 \geq\frac{1}{2}\left(\frac{\alpha+1}{\alpha-1}\right)^2$.
Let us denote $\beta = \frac{1}{2}\left(\frac{\alpha+1}{\alpha-1}\right)^2$. Then, the inequality holds if and only if $0<R<1$ and $(1-R\geq\beta$ or $1-R \leq-\beta)$. Using the distributive law, this can be simplified to 
\begin{equation}
    0<R<1 \quad\text{and}\quad \left((1-R\geq\beta)\quad\text{or} \quad (1-R \leq-\beta)\right).
\end{equation}
Since $\alpha>1$, we can deduce that $\beta>0$. Therefore, the case where $0<R<1$ and $1+\beta\leq R$ is not possible. Thus, the inequality holds if and only if $0<R<1$ and $(1-\beta\geq R)$.

Given the definitions of $R = \left(\dfrac{\kappa\delta}{\alpha-1}\right) \sqrt{\frac{\Omega}{E\pi}}$ and $\beta$, solving the quadratic equation $(1-\beta>R)$ for $\alpha$ yields
\begin{equation}
\label{eq:alpha_cond}
\left(\alpha \leq \left(3+t - \sqrt{t^2+6t+6}\right)\right) \quad\text{or} \quad\left(\alpha \geq\left(3+t + \sqrt{t^2+6t+6}\right)\right),
\end{equation}
where $t = \kappa\delta\sqrt{\frac{\Omega}{E\pi}}$ and $\alpha>1$. 
Therefore, we have shown that the upper bound for the IF-TEM MSE outperforms that of the classical ADC when \eqref{eq:alpha_cond} holds. Thereby indicating that the IF-TEM achieves superior MSE performance compared to the classical ADC in such instances This completed the theorem proof.
\bibliographystyle{IEEEtran}
\bibliography{main}
\end{document}